\documentclass[aps,pra,onecolumn,amsmath,amssymb,nofootinbib,superscriptaddress]{revtex4-2}
\usepackage[a4paper,left=25.4mm, right=20mm, bottom=20mm, top=20mm]{geometry} 
\usepackage{mathtext}
\usepackage{cancel}

\usepackage{setspace}
\usepackage{subcaption} 
\usepackage[T2A]{fontenc}
\usepackage[utf8x]{inputenc}
\usepackage[english]{babel}
\usepackage{indentfirst} 
\usepackage{cmap} 
\usepackage{textcomp} 
\usepackage{amssymb} 
\usepackage{amsmath} 
\usepackage{braket}
\usepackage{graphicx}  
\usepackage[small,raggedright]{titlesec}  
\usepackage{multirow}  
\usepackage{caption}
\usepackage[subfigure]{tocloft}
\usepackage{array} 
\usepackage{hyphenat}
\usepackage{float}
\usepackage{wrapfig} 
\usepackage{ccaption} 
\usepackage[usenames]{color} 
\usepackage{colortbl}
\usepackage{lastpage} %

\begin{document}
\title{Sideband fingerprints of antibunched light in cascaded quantum wave mixing}

\author{R. D. Ivanovskikh}
\affiliation{Dukhov Research Institute of Automatics (VNIIA), Moscow 127055, Russia}

\author{W. V. Pogosov}
\affiliation{Dukhov Research Institute of Automatics (VNIIA), Moscow 127055, Russia}
\affiliation{Moscow Institute of Physics and Technology, Dolgoprudny, 141700, Russia}
\affiliation{Institute for Theoretical and Applied Electrodynamics, Russian Academy of Sciences, Moscow 125412, Russia}

\author{A. A. Elistratov}
\affiliation{Dukhov Research Institute of Automatics (VNIIA), Moscow 127055, Russia}

\author{A. Yu. Dmitriev}
\affiliation{Moscow Institute of Physics and Technology, Dolgoprudny, 141700, Russia}
\affiliation{Kotelnikov Institute of Radioengineering and Electronics, Russian Academy of  Sciences, Moscow  125009, Russia }
\author{T. R. Sabirov}
\affiliation{Skolkovo Institute of Science and Technology, Nobel  St. 3, 143026, Moscow, Russia}
\affiliation{Moscow Institute of Physics and Technology, Dolgoprudny, 141700, Russia}

\author{A. V. Vasenin}
\affiliation{Moscow Institute of Physics and Technology, Dolgoprudny, 141700, Russia}

\author{S. A. Gunin}
\affiliation{Moscow Institute of Physics and Technology, Dolgoprudny, 141700, Russia}

\author{O. V. Astafiev}
\affiliation{Skolkovo Institute of Science and Technology, Nobel  St.  3, 143026, Moscow, Russia}
\affiliation{Kotelnikov Institute of Radioengineering and Electronics, Russian Academy  of  Sciences, Moscow 125009, Russia } 
\affiliation{Moscow Institute of Physics and Technology, Dolgoprudny, 141700, Russia}

\begin{abstract}
Quantum wave mixing on a single superconducting qubit produces a hierarchy of
coherent side peaks associated with elastic multiphoton scattering pathways. In
a cascaded source--probe geometry these pathways become sensitive to the photon
statistics of the radiation emitted by the source qubit. We develop an
analytical theory of this effect starting from the cascaded master equation in the weak-driving regime. In the coherent-filtering limit
when the decay rate of the probe qubit is less than that of the source, the standard coherent--coherent
wave-mixing hierarchy is recovered. In the opposite limit, side peaks associated with multiphoton
absorption from the antibunched source field are parametrically suppressed.
Numerical solutions confirm the analytical scaling laws. The resulting
sideband hierarchy provides a frequency-domain fingerprint of antibunched
itinerant microwave light.
\end{abstract}

\maketitle

\section{Introduction}

Wave mixing is a fundamental manifestation of optical nonlinearity. When two or
more fields interact with a nonlinear medium, the output spectrum contains new
components at frequencies given by integer combinations of the incident
frequencies, reflecting the corresponding multiphoton scattering processes and
energy conservation \cite{Shen1984,Boyd2008,Agrawal2007}. In conventional
nonlinear optics these processes are usually described in terms of nonlinear
susceptibilities of a macroscopic medium. A qualitatively different regime is
reached when the nonlinear element is reduced to a single quantum emitter. In
that case the sideband structure is governed by the quantum dynamics of an
individual two-level or few-level system, and the resulting spectrum can become
sensitive not only to the amplitudes and phases of the incident fields, but
also to their photon statistics.

Superconducting qubits coupled to one-dimensional microwave waveguides provide
a particularly suitable platform for this single-emitter nonlinear optics
regime
\cite{Astafiev2010,Roy2017,vanLoo2013,Hoi2013,Sathyamoorthy2014,Hofheinz2009,Peng2016,Zhou2020}.
Their strong effective nonlinearity, large radiative coupling, and high
spectral resolution make it possible to resolve coherent scattering components
generated by individual artificial atoms. They are also well suited for
studying nearly degenerate bichromatic driving, a regime that is natural in
circuit experiments and has a long history in atomic and optical nonlinear
spectroscopy \cite{Freedhoff1990,Agarwal1991,Ruyten1992}. In the context of
superconducting artificial atoms, wave mixing on a single qubit has been
studied under the name quantum wave mixing (QWM). Experiments first
demonstrated QWM for pulse trains \cite{Dmitriev2017}, then extended it to a
three-level artificial atom \cite{Decrinis2018}, and later to continuous
bichromatic driving of a two-level system \cite{Dmitriev2019}. In the latter
case, the scattered field contains a hierarchy of narrow coherent side peaks
whose amplitudes can be linked to spectral weights of elastic multiphoton
scattering pathways.

For the two incident frequencies used below, each elastic
wave-mixing pathway can be labeled by a pair of integers
$(n_{\rm s},n_{\rm pr})$ and produces a coherent output component at
\begin{equation}
    \omega_{n_{\rm s},n_{\rm pr}}
    =
    n_{\rm s}\omega_{\rm s}
    +
    n_{\rm pr}\omega_{\rm pr},
    \qquad
    n_{\rm s}+n_{\rm pr}=1.
\end{equation}
Positive integers count photons absorbed from the corresponding field,
whereas a negative integer represents stimulated emission into that
field. For the symmetrically detuned tones
$\omega_{\rm s}=\omega_{\rm tr}+\delta\omega$ and
$\omega_{\rm pr}=\omega_{\rm tr}-\delta\omega$, these processes generate
components at odd multiples of $\delta\omega$ around the qubit
transition. For example, the pathway
$(n_{\rm s},n_{\rm pr})=(-1,2)$ produces the component
$2\omega_{\rm pr}-\omega_{\rm s}
=\omega_{\rm tr}-3\delta\omega$.
The mixing originates from the nonlinear response of the two-level probe.
Because the probe response depends on its population, the two incident fields
modulate this response and generate new frequency components at integer
combinations of the two drive frequencies.

This pathway interpretation suggests a broader use of QWM. If a given side
peak is produced by a process involving a definite number of photons from each
incident field, then the side-peak hierarchy can serve as a probe of the
photon-number structure of the radiation driving the qubit. This idea was
developed theoretically for wave mixing between a coherent tone and
nonclassical incident fields, where photon statistics can impose selection
rules and suppress entire families of side peaks
\cite{paper1,Gardiner1991,RitschZoller1988,Breitenbach1997,Toyli2016}. From
this perspective, QWM is not only a spectroscopy of the nonlinear response of
the probe qubit, but also a spectroscopy of the quantum states of the propagating
field incident on it.

\begin{figure}[!h]\includegraphics[width=7cm]{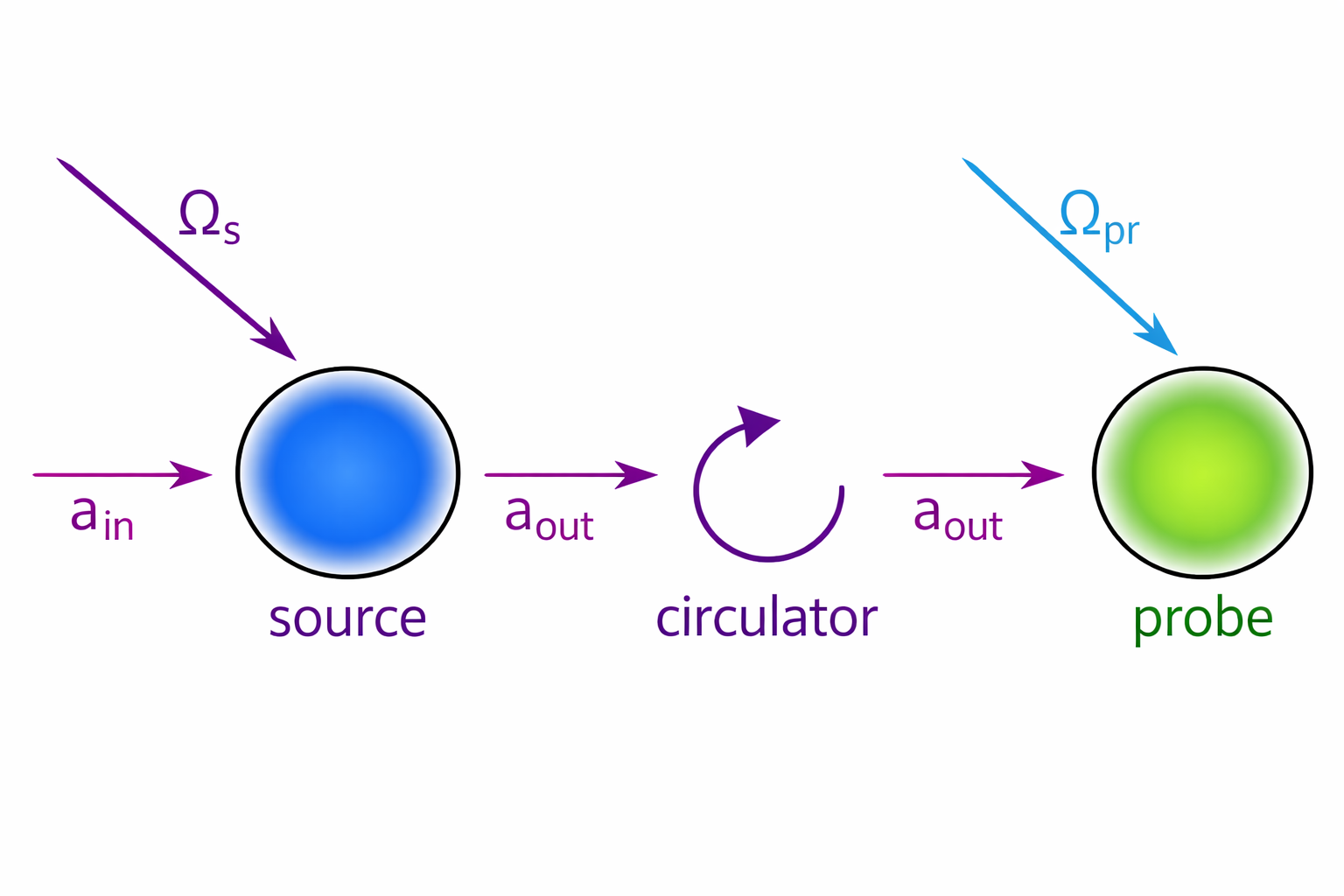} \caption{\raggedright Cascaded source--probe system. The circulator enforces unidirectional coupling: radiation emitted by the source drives the probe, while back-action from the probe to the source is suppressed. The input field $a_{\rm{in}}$ incident on the source is assumed to be vacuum. The coherent pump on the source is not included in $a_{\rm{in}}$.}\label{fig:cascaded} \end{figure}

A natural implementation of this idea is provided by a cascaded source--probe
geometry. In such a system, a driven source qubit emits resonance fluorescence
into a one-dimensional waveguide, and this propagating field irradiates a
second, probe qubit. The coupling is unidirectional: the source drives the
probe, while back-action from the probe to the source is suppressed
\cite{Astafiev2010,Gardiner1993,Gardiner1994}. In the microwave implementation, this directionality is provided by a circulator placed between the two qubits, as illustrated in Fig.~\ref{fig:cascaded}. The circulator routes the field emitted by the source toward the probe while suppressing propagation in the reverse direction. Thus, the unidirectional character of the source--probe coupling is imposed by the microwave network and does not rely on a particular relation between the radiative decay rates of the source and the probe. The probe can then be driven
simultaneously by the source radiation and by an externally applied coherent
tone. This geometry realizes wave mixing between a controlled coherent field
and a nonclassical field generated on chip by a single quantum emitter.

Recent work in this architecture demonstrated a pronounced suppression of QWM
side peaks associated with pathways requiring more than one photon from the
source field \cite{paper2}. This behavior is consistent with the antibunched
character of resonance fluorescence from a two-level system ~\cite{Kimble1977,filtered_statistics}. The effect was described by numerical simulations of the cascaded dynamics and interpreted in
terms of the photon statistics of the source radiation. What has remained
missing, however, is a compact analytical description that directly connects
the observed peak hierarchy with the parameters of the cascaded system and, in
particular, with the linewidth ratio of the source and probe qubits.

In this work we provide such an analytical description. Starting from the
cascaded master equation, we formulate the stationary response of the
source--probe system as a linear algebraic problem with two distinct parts. The
first part contains the decay rates and the unidirectional cascaded coupling,
whereas the second part is linear in the two coherent drive amplitudes. This
structure allows us to construct a systematic weak-drive Taylor expansion by
using a Neumann series for the inverse response operator. 

The resulting closed-form expressions give the leading Fourier components of
the probe coherence, and hence the amplitudes of the first QWM side peaks, in
terms of the source--probe coupling, the coherent drive amplitudes, and the
ratio $\gamma_{\rm{s}}/\gamma_{\rm{pr}}$. The formulas make explicit how the
same cascaded dynamics switches between two physically distinct regimes.
When $\gamma_{\rm{s}}\gg\gamma_{\rm{pr}}$, the probe effectively selects the
narrow coherent component of the source emission, and the known
coherent--coherent QWM results are recovered \cite{Dmitriev2019,paper1}. In
the opposite limit, $\gamma_{\rm{pr}}\gg\gamma_{\rm{s}}$, the probe is
sensitive to the antibunched fluorescence of the source, and higher-order side
peaks involving several photons from the source are parametrically suppressed.

The antibunching-dominated regime considered here complements, rather than reproduces, the strong-drive regime studied in Ref.~\cite{paper3}. In that work, a strongly and resonantly driven source together with a spectrally narrow probe selects the central Mollow component of the source fluorescence. Its normal and anomalous correlations act on the probe as an effective pair-correlated drive, giving a parity selection rule that suppresses QWM pathways containing an odd number of source-field photons. The present weak-drive regime is physically different: when the probe is broad on the scale of the source linewidth, it samples the antibunched resonance fluorescence of the source, and leading pathways containing two or more source-field photons are parametrically suppressed. The present suppression hierarchy is therefore not the parity rule of Ref.~\cite{paper3}. In particular, two-source-field pathways are suppressed here but belong to the allowed parity sector there, whereas three-source-field pathways are suppressed in both regimes for different physical reasons. The same cascaded architecture can thus exhibit distinct QWM fingerprints because the source excitation and the probe bandwidth select different statistical components of the emitted field.

These results provide a benchmark for numerical simulations of cascaded QWM and
a practical tool for interpreting experiments in which nonclassical microwave
radiation is characterized through probe's frequency-domain response. In this
sense, the hierarchy of QWM side peaks acts as a spectroscopic fingerprint of
antibunched itinerant light.

The paper is organized as follows. In Sec.~\ref{sec:eqs} we introduce the
cascaded source--probe model and the corresponding equations of motion. In
Sec.~\ref{sec:perturbations} we formulate the weak-drive Neumann expansion and
derive the leading QWM side-peak amplitudes. In
Sec.~\ref{sec:quantum and classical} we analyze the coherent-filtering and antibunching limits. In Sec.~\ref{sec:Discussion} we compare the
analytical predictions with numerical simulations and discuss the physical
origin of the peak suppression. Section~\ref{sec:Conclusion} summarizes the
results.

\section{Cascaded formalism and equations of motion}
\label{sec:eqs}

We consider a cascaded waveguide-QED system consisting of two two-level systems:
a source qubit, denoted by the index ${\rm s}$, and a probe qubit, denoted by
the index ${\rm pr}$. The unidirectional character of the coupling is
implemented in the standard cascaded setting \cite{Gardiner1993,Gardiner1994}:
radiation emitted by the source drives the probe, whereas back-action from the
probe to the source is suppressed. Experimentally, this directionality is implemented by a nonreciprocal
microwave element; in the setup considered here, a circulator placed between
the two qubits routes the source emission toward the probe and suppresses
reverse propagation as shown in Fig. \ref{fig:cascaded}. In contrast to a
Maxwell--Bloch description of two independent drives, the cascaded formalism
keeps the quantum source--probe correlations explicitly. The nonclassical
character of the radiation emitted by the source does not enter as an external
assumption, but is encoded in the equations of motion generated by the cascaded
master equation.

The origin of the cascaded coupling can be recalled within the standard
input--output description. The field emitted by the source satisfies
\begin{equation}
    a_{\rm out,s}(t)
    =
    a_{\rm in,s}(t)
    +
    \sqrt{\gamma_{\rm s}}\,\sigma_-^{\rm s}(t),
\end{equation}
where $a_{\rm in,s}$ is vacuum in the configuration considered here.
The circulator imposes the unidirectional boundary condition
\begin{equation}
    a_{\rm in,pr}(t)
    =
    \mu e^{i\phi}
    a_{\rm out,s}(t-\tau)
    +
    \sqrt{1-\mu^2}\,v(t),
\end{equation}
where $\mu$ is the field-amplitude transmission coefficient between the
qubits, $\phi$ is the propagation phase, $\tau$ is the propagation
delay, and $v(t)$ is the vacuum field associated with transmission
loss. The independently applied coherent probe tone is included in the
Hamiltonian below and is not part of $a_{\rm in,pr}$. In the Markovian,
delay-free limit, the delay is neglected and the constant phase can be
absorbed into the probe operators. The delay-free approximation requires the propagation time
$\tau=L/v_{\rm g}$ to be short compared with the relevant dynamical
timescales; over the corresponding frequency interval the
transmission amplitude and propagation phase can then be treated as
frequency independent. The Markov treatment also assumes that any stochastic field incident on
the source has a correlation time short compared with the relevant
source dynamics, or equivalently a bandwidth much larger than the
corresponding dynamical rates. This condition is automatically satisfied
for the vacuum input considered here. Eliminating the propagating field
then gives the standard cascaded master equation
\cite{Gardiner1993,Gardiner1994}
\begin{equation}\label{app::master_eq Gardiner}
\begin{split}
\dfrac{d\rho}{dt}=&\dfrac{i}{\hbar}[\rho, H_{\rm{sys}}]
-\sqrt{\gamma_{\rm{pr}}\gamma_{\rm{s}}}\mu
\left([\sigma^{\rm{pr}}_{+},\sigma^{\rm{s}}_{-}\rho]
+[\rho\sigma^{\rm{s}}_{+},\sigma^{\rm{pr}}_{-}]\right)
+\hat{L}_{1}\rho+\hat{L}_{2}\rho .
\end{split}
\end{equation}
Here $\rho$ is the density matrix of the two-qubit system,
$\sigma_{\pm}^{\rm{s,pr}}$ are the raising and lowering Pauli operators in the
source and probe subspaces, $\gamma_{\rm{s}}$ and $\gamma_{\rm{pr}}$ are their
radiative decay rates. The Hermitian Hamiltonian is
\[
\begin{split}
H_{\rm{sys}}=&\dfrac{1}{2}\hbar\omega_{\rm{tr1}}\sigma_{z}^{\rm{s}}
+\dfrac{1}{2}\hbar\omega_{\rm{tr2}}\sigma_{z}^{\rm{pr}}\\
&+
\left(\Omega_{\rm{s}}e^{-i\omega_{\rm{s}}t}\sigma_{+}^{\rm{s}}
+\text{h.c.} \right)
+
\left(\Omega_{\rm{pr}}e^{-i\omega_{\rm{pr}}t}\sigma_{+}^{\rm{pr}}
+\text{h.c.} \right).
\end{split}
\]
that describes the unitary evolution of the system and includes the coherent drive, where $\Omega_{\rm{s}}$, $\omega_{\rm{s}}$, $\omega_{\rm{tr1}}$  and $\Omega_{\rm{pr}}$,
$\omega_{\rm{pr}}$, $\omega_{\rm{tr2}}$ --- the Rabi frequency, drive frequency and transition frequency of the source qubit and probe qubit, respectively.
We assume, for simplicity, that
$\omega_{\rm{tr1}}=\omega_{\rm{tr2}}=\omega_{\rm{tr}}$ and choose the two drive
frequencies as
\[
\omega_{\rm{pr}}=\omega_{\rm{tr}}-\delta\omega,
\qquad
\omega_{\rm{s}}=\omega_{\rm{tr}}+\delta\omega .
\]
In the analytical derivation below, we focus on the near-resonance regime
$\delta\omega\ll\gamma_{\rm{s}},\gamma_{\rm{pr}}$, where the slowly varying phase $\delta\omega t$ on the scales $\gamma_{\rm{s}}^{-1},\gamma_{\rm{pr}}^{-1}$ provides the validity of the stationary approximation that will be used later. The dissipators
are
\[
\begin{split}
\hat{L}_{1}\rho&=\dfrac{1}{2}\gamma_{\rm{s}}
\left(2\sigma_{-}^{\rm{s}}\rho\sigma_{+}^{\rm{s}}
-\sigma_{+}^{\rm{s}}\sigma_{-}^{\rm{s}}\rho
-\rho\sigma_{+}^{\rm{s}}\sigma_{-}^{\rm{s}}\right),\\
\hat{L}_{2}\rho&=
\dfrac{1}{2}\gamma_{\rm{pr}}
\left(2\sigma_{-}^{\rm{pr}}\rho\sigma_{+}^{\rm{pr}}
-\sigma_{+}^{\rm{pr}}\sigma_{-}^{\rm{pr}}\rho
-\rho\sigma_{+}^{\rm{pr}}\sigma_{-}^{\rm{pr}}\right).
\end{split}
\]

The asymmetric cross term in Eq.~(\ref{app::master_eq Gardiner})
transfers source-field fluctuations to the probe dynamics while leaving
the source evolution independent of the probe, as required for a
unidirectional cascade.

We work in a frame rotating at $\omega_{\rm tr}$ and introduce 
dimensionless time $\tau=\gamma_{\rm{pr}}t$. The dimensionless cascaded
coupling is
\begin{equation}\label{alpha}
\alpha=\mu\sqrt{\dfrac{\gamma_{\rm{s}}}{\gamma_{\rm{pr}}}}.
\end{equation}
From Eq.~(\ref{app::master_eq Gardiner}) one obtains the following set of
equations for the probe coherence, probe inversion, and source--probe
correlators:
\begin{equation}\label{sigma_m^p}
\dfrac{\partial\braket{{\sigma}_{-}^{\rm{pr}}}}{\partial{\tau}}
=
\frac{\Omega_{\rm{pr}}}{\gamma_{\rm{pr}}}
\braket{\sigma_{z}^{\rm{pr}}} e^{- i \delta\omega t}
+\alpha \braket{\sigma_{-}^{\rm{s}}\sigma_{z}^{\rm{pr}}}
-\frac{\braket{\sigma_{-}^{\rm{pr}}}}{2},
\end{equation}
\begin{equation}\label{sigma_z^p}
\begin{split}
\dfrac{\partial\braket{{\sigma}_{z}^{\rm{pr}}}}{\partial\tau}
=&-
\left(\frac{2 \Omega_{\rm{pr}}}{\gamma_{\rm{pr}}}
\braket{\sigma_{+}^{\rm{pr}}} e^{- i \delta\omega t}
+\frac{2\overline{\Omega_{\rm{pr}}}}{\gamma_{\rm{pr}}}
\braket{\sigma_{-}^{\rm{pr}}} e^{i \delta\omega t} \right)-2 \alpha
\left( \braket{\sigma_{+}^{\rm{s}}\sigma_{-}^{\rm{pr}}}
+\braket{\sigma_{-}^{\rm{s}}\sigma_{+}^{\rm{pr}}}\right)
-\braket{\sigma_{z}^{\rm{pr}}}-1,
\end{split}
\end{equation}
\begin{equation}\label{sigma_mp}
\begin{split}
\dfrac{\partial\braket{{\sigma_{-}^{\rm{s}}\sigma_{+}^{\rm{pr}}}}}{\partial\tau}
=&
\frac{\Omega_{\rm{s}}}{\gamma_{\rm{pr}}}
\braket{\sigma_{z}^{\rm{s}}\sigma_{+}^{\rm{pr}}} e^{i \delta\omega t}
+
\dfrac{\overline{\Omega_{\rm{pr}}}}{\gamma_{\rm{pr}}}
\braket{\sigma_{-}^{\rm{s}}\sigma_{z}^{\rm{pr}}} e^{i \delta\omega t}+\alpha\left(
\frac{ \braket{\sigma_{z}^{\rm{pr}}}}{2}
+\frac{\braket{\sigma_{z}^{\rm{s}}\sigma_{z}^{\rm{pr}}}}{2}\right)
-\braket{\sigma_{-}^{\rm{s}}\sigma_{+}^{\rm{pr}}}
\left( \frac{\gamma_{\rm{s}}}{2 \gamma_{\rm{pr}}}+\frac{1}{2}\right),
\end{split}
\end{equation}
\begin{equation}\label{sigma_pp}
\begin{split}
\dfrac{\partial\braket{{\sigma_{+}^{\rm{s}}\sigma_{+}^{\rm{pr}}}}}{\partial\tau}
=&
-\braket{\sigma_{+}^{\rm{s}}\sigma_{+}^{\rm{pr}}}
\left( \frac{\gamma_{\rm{s}}}{2 \gamma_{\rm{pr}}}+\frac{1}{2}\right)
+
\frac{\overline{\Omega_{\rm{pr}}}}{\gamma_{\rm{pr}}}
\braket{\sigma_{+}^{\rm{s}}\sigma_{z}^{\rm{pr}}} e^{i \delta\omega t}+
\frac{\overline{\Omega_{\rm{s}}}}{\gamma_{\rm{pr}}}
\braket{\sigma_{z}^{\rm{s}}\sigma_{+}^{\rm{pr}}} e^{- i \delta\omega t},
\end{split}
\end{equation}
\begin{equation}\label{sigma_pz}
\begin{split}
\dfrac{\partial\braket{{\sigma_{+}^{\rm{s}}\sigma_{z}^{\rm{pr}}}}}{\partial\tau}
=&-
\left(
\frac{2 \Omega_{\rm{pr}}}{\gamma_{\rm{pr}}}
\braket{\sigma_{+}^{\rm{s}}\sigma_{+}^{\rm{pr}}} e^{- i \delta\omega t}
+\frac{2\overline{\Omega_{\rm{pr}}}}{\gamma_{\rm{pr}}}
\braket{\sigma_{+}^{\rm{s}}\sigma_{-}^{\rm{pr}}} e^{i \delta\omega t}
\right)+
\frac{\overline{\Omega_{\rm{s}}}}{\gamma_{\rm{pr}}}
\braket{\sigma_{z}^{\rm{s}}\sigma_{z}^{\rm{pr}}} e^{- i \delta\omega t}
-\alpha \braket{\sigma_{+}^{\rm{pr}}}
-\alpha \braket{\sigma_{z}^{\rm{s}}\sigma_{+}^{\rm{pr}}}\\
&-
\braket{\sigma_{+}^{\rm{s}}\sigma_{z}^{\rm{pr}}}
\left( \frac{\gamma_{\rm{s}}}{2 \gamma_{\rm{pr}}}+1\right)
-\braket{\sigma_{+}^{\rm{s}}},
\end{split}
\end{equation}
\begin{equation}\label{sigma_zz}
\begin{split}
\dfrac{\partial\braket{{\sigma_{z}^{\rm{s}}\sigma_{z}^{\rm{pr}}}}}{\partial\tau}
=&-
\left(
\frac{2 \Omega_{\rm{pr}}}{\gamma_{\rm{pr}}}
\braket{\sigma_{z}^{\rm{s}}\sigma_{+}^{\rm{pr}}} e^{- i \delta\omega t}
+
\frac{2\overline{\Omega_{\rm{pr}}}}{\gamma_{\rm{pr}}}
\braket{\sigma_{z}^{\rm{s}}\sigma_{-}^{\rm{pr}}} e^{i \delta\omega t}
\right)-
\left(
\frac{2 \Omega_{\rm{s}}}{\gamma_{\rm{pr}}}
\braket{\sigma_{+}^{\rm{s}}\sigma_{z}^{\rm{pr}}} e^{i \delta\omega t}
+
\frac{2\overline{\Omega_{\rm{s}}}}{\gamma_{\rm{pr}}}
\braket{\sigma_{-}^{\rm{s}}\sigma_{z}^{\rm{pr}}} e^{- i \delta\omega t}
\right)\\
&+
2 \alpha
\left(\braket{\sigma_{+}^{\rm{s}}\sigma_{-}^{\rm{pr}}}
+\braket{\sigma_{-}^{\rm{s}}\sigma_{+}^{\rm{pr}}}\right)
-\braket{\sigma_{z}^{\rm{pr}}}\frac{ \gamma_{\rm{s}}}{\gamma_{\rm{pr}}}-
\braket{\sigma_{z}^{\rm{s}}\sigma_{z}^{\rm{pr}}}
\left( \frac{\gamma_{\rm{s}}}{\gamma_{\rm{pr}}}+1\right)
-\braket{\sigma_{z}^{\rm{s}}},
\end{split}
\end{equation}
\begin{equation}\label{sigma_zm}
\begin{split}
\dfrac{\partial\braket{{\sigma_{z}^{\rm{s}}\sigma_{-}^{\rm{pr}}}}}{\partial\tau}
=&
\frac{\Omega_{\rm{pr}}}{\gamma_{\rm{pr}}}
\braket{\sigma_{z}^{\rm{s}}\sigma_{z}^{\rm{pr}}} e^{- i \delta\omega t}-
\left(
\frac{2 \Omega_{\rm{s}}}{\gamma_{\rm{pr}}}
\braket{\sigma_{+}^{\rm{s}}\sigma_{-}^{\rm{pr}}} e^{i \delta\omega t}
+
\frac{2 \overline{\Omega_{\rm{s}}}}{\gamma_{\rm{pr}}}
\braket{\sigma_{-}^{\rm{s}}\sigma_{-}^{\rm{pr}}} e^{- i \delta\omega t}
\right)\\
&-\alpha \braket{\sigma_{-}^{\rm{s}}\sigma_{z}^{\rm{pr}}}
-\braket{\sigma_{-}^{\rm{pr}}}\frac{ \gamma_{\rm{s}}}{\gamma_{\rm{pr}}}
-\braket{\sigma_{z}^{\rm{s}}\sigma_{-}^{\rm{pr}}}
\left( \frac{\gamma_{\rm{s}}}{\gamma_{\rm{pr}}}+\frac{1}{2}\right).
\end{split}
\end{equation}

The source subsystem is decoupled from the probe, as required by the cascaded
construction. Its equations are the usual Maxwell--Bloch equations,
\[
\frac{\partial\braket{{\sigma}_{-}^{\rm{s}}}}{\partial\tau}
=
\frac{\Omega_{s} \braket{\sigma_{z}^{\rm{s}}} e^{i \delta\omega t}}
{\gamma_{\rm{pr}}}
-
\frac{\gamma_{s}}{2\gamma_{\rm{pr}}}
\braket{\sigma_{-}^{\rm{s}}},
\]
\[
\frac{\partial\braket{{\sigma}_{z}^{\rm{s}}}}{\partial\tau}
=
-\frac{2 \Omega_{s}}{\gamma_{\rm{pr}}}
\braket{\sigma_{+}^{\rm{s}}} e^{i \delta\omega t}
-\frac{2 \overline{\Omega_{s}}}{\gamma_{\rm{pr}}}
\braket{\sigma_{-}^{\rm{s}}} e^{- i \delta\omega t}
-\frac{\gamma_{s}}{\gamma_{\rm{pr}}}
\braket{\sigma_{z}^{\rm{s}}}
-\frac{\gamma_{s}}{\gamma_{\rm{pr}}}.
\]
Their stationary solution is
\begin{equation}\label{sigma_z^s}
\braket{\sigma_{z}^{\rm{s}}}
=
-\dfrac{1}{1+\dfrac{8|\Omega_{s}|^2}{\gamma_{s}^{2}}},
\end{equation}
\begin{equation}\label{sigma_m^s}
\braket{\sigma_{-}^{\rm{s}}}
=
-\dfrac{2\Omega_{s}e^{i\delta\omega t}}
{\gamma_s\left(1+\dfrac{8|\Omega_{s}|^2}{\gamma_{s}^{2}}\right)}.
\end{equation}
These source solutions enter the remaining equations as known inhomogeneous
terms.

\section{Weak-drive Taylor expansion of the cascaded response}
\label{sec:perturbations}

The equations derived above form a closed linear system for the probe variables
and source--probe correlators once the source averages
(\ref{sigma_z^s}) and (\ref{sigma_m^s}) are substituted. The main point of the
analytical treatment is to organize this linear system so that the linewidth
ratio and the cascaded coupling are kept exactly, whereas the response is
expanded only in the two weak coherent drives.

We introduce
\begin{equation}
r=\frac{\gamma_{\rm{s}}}{\gamma_{\rm{pr}}},
\qquad
\eta=e^{i\delta\omega t},
\end{equation}
and define the four drive amplitudes
\begin{equation}\label{drive variables}
p_-=\frac{\Omega_{\rm{pr}}}{\gamma_{\rm{pr}}}\eta^{-1},
\qquad
p_+=\frac{\overline{\Omega_{\rm{pr}}}}{\gamma_{\rm{pr}}}\eta,
\end{equation}
\begin{equation}
s_+=\frac{\Omega_{\rm{s}}}{\gamma_{\rm{pr}}}\eta,
\qquad
s_-=\frac{\overline{\Omega_{\rm{s}}}}{\gamma_{\rm{pr}}}\eta^{-1}.
\end{equation}
Here \(p_\pm\) denote the probe drive components at \(\mp\delta\omega\), while
\(s_\pm\) denote the source drive components at \(\pm\delta\omega\). The weak
driving regime is
\begin{equation}\label{weak-drive cond1}
\dfrac{|\Omega_{\rm{pr}}|}{\gamma_{\rm{pr}}}=|p_{\pm}|\ll1,\ \ \ \ \ \ \dfrac{|\Omega_{\rm{s}}|}{\gamma_{\rm{s}}}=|s_{\pm}|/r\ll1.
\end{equation}

We collect the dynamical variables in the vector
\begin{equation}\label{vector X}
\begin{split}
\vec X=&
\bigl(
\braket{\sigma_{-}^{\rm{pr}}},
\braket{\sigma_{-}^{\rm{s}}\sigma_{z}^{\rm{pr}}},
\braket{\sigma_{z}^{\rm{s}}\sigma_{-}^{\rm{pr}}},
\braket{\sigma_{+}^{\rm{pr}}},
\braket{\sigma_{+}^{\rm{s}}\sigma_{z}^{\rm{pr}}},
\braket{\sigma_{z}^{\rm{s}}\sigma_{+}^{\rm{pr}}},
\\
&\braket{\sigma_{-}^{\rm{s}}\sigma_{-}^{\rm{pr}}},
\braket{\sigma_{+}^{\rm{s}}\sigma_{+}^{\rm{pr}}},
\braket{\sigma_{z}^{\rm{pr}}},
\braket{\sigma_{+}^{\rm{s}}\sigma_{-}^{\rm{pr}}},
\braket{\sigma_{-}^{\rm{s}}\sigma_{+}^{\rm{pr}}},
\braket{\sigma_{z}^{\rm{s}}\sigma_{z}^{\rm{pr}}}
\bigr)^T .
\end{split}
\end{equation}
In the stationary approximation $\delta\omega\ll\gamma_{\rm{s}},\gamma_{\rm{pr}}$ we set the derrivatives on the left-hand side of Eqs.(\ref{sigma_m^p})-(\ref{sigma_zm}) to zero and the equations can be written as
\begin{equation}\label{matrix form}
0=(\hat A+\hat\Omega)\vec X+\vec b .
\end{equation}
The matrix \(\hat A\) contains all terms independent of the coherent drives,
including the decay rates and the cascaded coupling \(\alpha\). The matrix
\(\hat\Omega\) is linear in \(p_\pm\) and \(s_\pm\). The vector \(\vec b\)
contains the inhomogeneous terms generated by the ground-state contribution and
by the source averages. Explicit expressions for \(\hat A\), \(\hat\Omega\),
and \(\vec b\) are given in Appendix~\ref{app:matrices}.

The formal stationary solution is
\begin{equation}
\vec X=-(\hat A+\hat\Omega)^{-1}\vec b .
\end{equation}
Since \(\hat\Omega\) is linear in the weak drive amplitudes, the inverse
operator can be expanded as a Neumann series,
\begin{equation}\label{Neumann}
(\hat A+\hat\Omega)^{-1}
=
\left(\hat 1+\hat A^{-1}\hat\Omega\right)^{-1}\hat A^{-1}
=
\left[
\hat 1-\hat A^{-1}\hat\Omega
+\left(\hat A^{-1}\hat\Omega\right)^2-\ldots
\right]\hat A^{-1}.
\end{equation}
Thus
\begin{equation}\label{formal solution Neumann}
\vec X
=
-\left[
\hat 1-\hat A^{-1}\hat\Omega
+\left(\hat A^{-1}\hat\Omega\right)^2-\ldots
\right]\hat A^{-1}\vec b .
\end{equation}
This representation is the central technical step: the linewidth ratio \(r\)
and the cascaded coupling \(\alpha\) are retained in \(\hat A^{-1}\), while the expansion parameters are the source and probe drive strengths,
$\Omega_{\rm s}$ and $\Omega_{\rm pr}$.

The source-dependent part of \(\vec b\) is also expanded in powers of the source
drive. With
\begin{equation}
F_s=\frac{1}{1+8s_+s_-/r^2}
=
1-\frac{8s_+s_-}{r^2}
+\frac{64s_+^2s_-^2}{r^4}
+O(|s|^6),
\end{equation}
the source averages become
\begin{equation}
-\braket{\sigma_-^{\rm s}}=\frac{2s_+}{r}F_s,
\qquad
-\braket{\sigma_+^{\rm s}}=\frac{2s_-}{r}F_s,
\qquad
-\braket{\sigma_z^{\rm s}}=F_s .
\end{equation}
We write
\begin{equation}
\vec X=\sum_{N=0}^{\infty}\vec X^{[N]},
\qquad
\vec b=\sum_{N=0}^{\infty}\vec b^{[N]},
\end{equation}
where the superscript \([N]\) denotes the total order in
\(p_\pm,s_\pm\). Equation~(\ref{matrix form}) then gives the recursion
\begin{equation}\label{recursion}
\vec X^{[N]}
=
-\hat A^{-1}
\left(
\hat\Omega\vec X^{[N-1]}+\vec b^{[N]}
\right),
\qquad N\geq 1,
\end{equation}
with
\begin{equation}
\vec X^{[0]}=-\hat A^{-1}\vec b^{[0]}.
\end{equation}
This recursion generates the weak-drive Taylor expansion of the full cascaded
solution. It also provides a simple selection rule. A monomial
\[
p_-^a p_+^b s_+^c s_-^d
\]
contributes to the Fourier component at
\begin{equation}\label{ndomega}
n\delta\omega=(-a+b+c-d)\delta\omega .
\end{equation}

Equation~(\ref{ndomega}) has a direct interpretation in terms of multiphoton scattering. For a monomial $p_-^a p_+^b s_+^c s_-^d$, define the signed numbers of photons participating from the source and probe fields as \begin{equation} n_{\rm{s}}=c-d, \qquad n_{\rm{pr}}=a-b . \end{equation} Positive values correspond to net absorption from the corresponding incident field, whereas negative values correspond to stimulated emission into that field. The coherent output component generated by such a process has the frequency 
\begin{equation} 
\omega_{\rm{out}} = n_{\rm{s}}\omega_{\rm{s}} + n_{\rm{pr}}\omega_{\rm{pr}},
\end{equation} 
for the sideband-producing terms in the probe coherence, $n_{\rm{s}}+n_{\rm{pr}}=1$, which is in agreement with Eq.~(\ref{ndomega}). For example, the monomial $p_-^2s_-$ corresponds to $(n_{\rm{s}},n_{\rm{pr}})=(-1,2)$ and generates the $-3\delta\omega$ component at $2\omega_{\rm{pr}}-\omega_{\rm{s}}$.

Finally, from the input--output relation the coherent field radiated by the probe into the downstream detection channel is proportional to $\sqrt{\gamma_{\rm{pr}}}\langle\sigma_-^{\rm{pr}}\rangle$. Thus, the Fourier components of $\langle\sigma_-^{\rm{pr}}\rangle$ calculated below are directly proportional to the coherent QWM side-peak amplitudes measured in the output spectrum.

We now apply Eq.~(\ref{recursion}) to the first component of \(\vec X\),
\[
X_1=\braket{\sigma_-^{\rm pr}}.
\]
To first order in the drives one obtains
\begin{equation}\label{x1 first ps}
X_1^{[1]}=-2p_-+\frac{4\alpha}{r}s_+ .
\end{equation}
In physical variables this gives the two leading coherent components
\begin{equation}\label{C minus one}
\braket{\sigma_-^{\rm pr}}_{-\delta\omega}
=
-\frac{2\Omega_{\rm pr}}{\gamma_{\rm pr}}e^{-i\delta\omega t},
\end{equation}
\begin{equation}\label{+1 peak}
\braket{\sigma_-^{\rm pr}}_{+\delta\omega}
=
\frac{4\alpha\Omega_{\rm s}}{\gamma_{\rm s}}e^{i\delta\omega t}
=
\mu\sqrt{\frac{\gamma_{\rm s}}{\gamma_{\rm pr}}}
\frac{4\Omega_{\rm s}}{\gamma_{\rm s}}e^{i\delta\omega t}.
\end{equation}

Equation~(\ref{weak-drive cond1}) also shows that weak excitation of the source itself
does not, in general, guarantee that the field emitted by the source
constitutes a weak drive for the probe. The source-induced contribution
to the probe coherence is controlled by the dimensionless combination
$\alpha|\Omega_{\rm{s}}|/\gamma_{\rm{s}}$. Therefore, the weak-drive
conditions~(\ref{weak-drive cond1}) should be supplemented by
\[
    \frac{\alpha |s_\pm|}{r}\ll1
\]

Therefore, the summarized conditions for the weak-drive expansion are
\begin{equation}\label{weak drive cond summ}
\dfrac{|\Omega_{\rm{pr}}|}{\gamma_{\rm{pr}}}\ll1,\ \ \dfrac{|\Omega_{\rm{s}}|}{\gamma_{\rm{s}}}\ll1,\ \ \alpha\frac{|\Omega_{\rm{s}}|}{\gamma_{\rm{s}}}
    \ll1,
\end{equation}
and the condition for the stationary approximation
\[
\delta\omega\ll\gamma_{\rm{s}}, \gamma_{\rm{pr}}.
\]

The full third-order correction to the probe coherence is
\begin{equation}\label{x1 third full}
\begin{split}
X_1^{[3]}={}&
16p_-^2p_+
-\frac{32\alpha}{r}p_-^2s_-
-\frac{64\alpha}{r}p_-p_+s_+
+\frac{128\alpha^2}{r^2}p_-s_+s_-\\
&+
\frac{64\alpha^2}{r(r+1)}p_+s_+^2
-\frac{32\alpha(4\alpha^2r+r+1)}
{r^3(r+1)}
s_+^2s_- .
\end{split}
\end{equation}
The terms proportional to \(p_-^2p_+\), \(p_-p_+s_+\),
\(p_-s_+s_-\), and \(s_+^2s_-\) provide third-order corrections to the
components at \(\pm\delta\omega\). The new QWM side peaks at third order are generated by
\(p_-^2s_-\) and \(p_+s_+^2\). Therefore
\begin{equation}\label{-3 peak}
\braket{\sigma_-^{\rm pr}}_{-3\delta\omega}
=
-32\alpha
\frac{\Omega_{\rm pr}^2\overline{\Omega_{\rm s}}}
{\gamma_{\rm s}\gamma_{\rm pr}^2}
e^{-3i\delta\omega t}.
\end{equation}
The opposite side peak is
\begin{equation}\label{+3 peak}
\braket{\sigma_-^{\rm pr}}_{+3\delta\omega}
=
64\alpha^2
\frac{\Omega_{\rm s}^2\overline{\Omega_{\rm pr}}}
{\gamma_{\rm s}\gamma_{\rm pr}
(\gamma_{\rm s}+\gamma_{\rm pr})}
e^{3i\delta\omega t}.
\end{equation}

The fifth-order calculation proceeds in exactly the same way. Although the full
expression for \(X_1^{[5]}\) is lengthy, the sideband-selective terms have a
compact form. The component at \(-5\delta\omega\) is generated by
\(p_-^3s_-^2\):
\begin{equation}\label{-5 peak}
\braket{\sigma_-^{\rm pr}}_{-5\delta\omega}
=
-\alpha^2
\frac{
(512\gamma_{\rm s}+768\gamma_{\rm pr})
\Omega_{\rm pr}^3\overline{\Omega_{\rm s}}^2
}
{
\gamma_{\rm s}\gamma_{\rm pr}^3
(\gamma_{\rm s}+\gamma_{\rm pr})^2
}
e^{-5i\delta\omega t}.
\end{equation}
The component at \(+5\delta\omega\) is generated by \(p_+^2s_+^3\):
\begin{equation}\label{+5 peak}
\braket{\sigma_-^{\rm pr}}_{+5\delta\omega}
=
\alpha^3
\frac{
(1024\gamma_{\rm s}+2560\gamma_{\rm pr})
\Omega_{\rm s}^3\overline{\Omega_{\rm pr}}^2
}
{
\gamma_{\rm s}\gamma_{\rm pr}^2
\left(
\gamma_{\rm s}^{3}
+4\gamma_{\rm s}^{2}\gamma_{\rm pr}
+5\gamma_{\rm s}\gamma_{\rm pr}^{2}
+2\gamma_{\rm pr}^{3}
\right)
}
e^{5i\delta\omega t}.
\end{equation}
Equivalently, the last denominator can be written as
\[
\gamma_{\rm s}\gamma_{\rm pr}^2
(\gamma_{\rm s}+\gamma_{\rm pr})^2
(\gamma_{\rm s}+2\gamma_{\rm pr}).
\]

Equations~(\ref{+1 peak})--(\ref{+5 peak}) are the main analytical result of
the paper. They provide the leading QWM side-peak amplitudes as closed
functions of the two drive amplitudes, the cascaded coupling, and the linewidth
ratio. The derivation also makes transparent which monomial in the two incident
fields generates each side peak: \(-3\delta\omega\) is associated with
\(p_-^2s_-\), \(+3\delta\omega\) with \(p_+s_+^2\),
\(-5\delta\omega\) with \(p_-^3s_-^2\), and \(+5\delta\omega\) with
\(p_+^2s_+^3\).

\section{Coherent-filtering and antibunching limits}
\label{sec:quantum and classical}

The closed-form amplitudes derived above become particularly transparent in
two opposite limits of the linewidth ratio. These limits separate two physical
regimes. In the first one the source acts effectively as a coherent tone for
the probe. In the second one the probe is sensitive to the 
antibunched character of the source fluorescence.

\subsection{Coherent-filtering limit $\gamma_{\rm s}\gg\gamma_{\rm pr}$}
\label{subsec:coherent}

When \(\gamma_{\rm s}\gg\gamma_{\rm pr}\), the probe resolves only the narrow
coherent component of the field emitted by the source, while the incoherent
fluorescence background is filtered out. The cascaded problem then reduces to
standard QWM of two coherent waves incident on the probe qubit.

For comparison, consider a single qubit with decay rate \(\gamma\) driven by
two coherent tones with slowly varying amplitudes
\(\Omega_1e^{-i\delta\omega t}\) and
\(\Omega_2e^{i\delta\omega t}\). In the same weak-drive approximation, the
leading coherent wave-mixing amplitudes are
\begin{equation}\label{coh+coh +-1 delta}
\braket{\sigma_{-}}_{+1\delta\omega}^{\rm{coh+coh}}=\dfrac{2\Omega_{2}}{\gamma},\ \  \braket{\sigma_{-}}_{-1\delta\omega}^{\rm{coh+coh}}=\dfrac{2\Omega_{1}}{\gamma},
\end{equation}
\begin{equation}\label{coh+coh +-3 delta}
\braket{\sigma_{-}}_{+3\delta\omega}^{\rm{coh+coh}}=\dfrac{16\Omega_{1}\Omega_{2}^{2}}{\gamma^{3}},\ \  \braket{\sigma_{-}}_{-3\delta\omega}^{\rm{coh+coh}}=\dfrac{16\Omega_{2}\Omega_{1}^{2}}{\gamma^{3}},
\end{equation}
\begin{equation}\label{coh+coh +-5 delta}
\braket{\sigma_{-}}_{+5\delta\omega}^{\text{coh+coh}}=\frac{128 \Omega_{2}^{3} \Omega_{1}^{2}}{\gamma^{5}},\ \ \braket{\sigma_{-}}_{-5\delta\omega}^{\text{coh+coh}}=\frac{128 \Omega_{1}^{3} \Omega_{2}^{2}}{\gamma^{5}}.
\end{equation}
These expressions are the weak-drive expansion of coherent--coherent QWM
\cite{Dmitriev2019,paper1}.

In the cascaded problem, taking the limit
\(\gamma_{\rm s}\gg\gamma_{\rm pr}\) in
Eqs.~(\ref{+1 peak})--(\ref{+5 peak}) reproduces
Eqs.~(\ref{coh+coh +-1 delta})--(\ref{coh+coh +-5 delta}) with
\begin{equation}\label{effective Rabis}
\gamma=\gamma_{\rm pr},
\qquad
\Omega_1=\Omega_{\rm pr},
\qquad
\Omega_2^{\rm eff}
=
-2\mu\Omega_{\rm s}
\sqrt{\frac{\gamma_{\rm pr}}{\gamma_{\rm s}}}.
\end{equation}
The sign of \(\Omega_2^{\rm eff}\) is a phase convention inherited from the
cascaded coupling term and has no effect on the measured side-peak intensities.
Thus the Neumann expansion passes an important consistency check: in the
coherent-filtering limit the cascaded source reduces to an effective coherent
drive.

It is often useful to express the Rabi amplitudes through drive voltages,
\begin{equation}\label{voltages def}
\Omega_{\rm pr}=\sqrt{\gamma_{\rm pr}}\varepsilon_{\rm pr},
\qquad
\Omega_{\rm s}=\sqrt{\gamma_{\rm s}}\varepsilon_{\rm s}.
\end{equation}
Then \(\Omega_2^{\rm eff}=-2\mu\sqrt{\gamma_{\rm pr}}\varepsilon_{\rm s}\),
so the coherent-limit side-peak amplitudes do not depend on
\(\gamma_{\rm s}\) at fixed source voltage.

\subsection{Antibunching limit $\gamma_{\rm pr}\gg\gamma_{\rm s}$}
\label{subsec:single-photon}

In the opposite limit, \(\gamma_{\rm pr}\gg\gamma_{\rm s}\), the probe is
broadband on the scale of the source linewidth and is sensitive to the full
resonance fluorescence emitted by the source. This radiation is antibunched for
a two-level source \cite{Astafiev2010}, and the wave-mixing pathways involving
several source photons are suppressed.

Taking the limit \(\gamma_{\rm pr}\gg\gamma_{\rm s}\) in the analytical
amplitudes gives
\begin{equation}\label{1-ph +3 peak}
\braket{\sigma_-^{\rm pr}}^{\rm ab}_{+3\delta\omega}
=
\mu^2\frac{\gamma_{\rm s}}{\gamma_{\rm pr}}
\frac{
64\Omega_{\rm s}^{2}\overline{\Omega_{\rm pr}}
}
{
\gamma_{\rm s}\gamma_{\rm pr}^{2}
}
e^{3i\delta\omega t},
\end{equation}
\begin{equation}\label{1-ph -5 peak}
\braket{\sigma_-^{\rm pr}}^{\rm ab}_{-5\delta\omega}
=
-\mu^2\frac{\gamma_{\rm s}}{\gamma_{\rm pr}}
\frac{
768\Omega_{\rm pr}^{3}\overline{\Omega_{\rm s}}^{\,2}
}
{
\gamma_{\rm s}\gamma_{\rm pr}^{4}
}
e^{-5i\delta\omega t},
\end{equation}
and
\begin{equation}\label{1-ph +5 peak}
\braket{\sigma_-^{\rm pr}}^{\rm ab}_{+5\delta\omega}
=
\mu^3
\left(\frac{\gamma_{\rm s}}{\gamma_{\rm pr}}\right)^{3/2}
\frac{
1280\Omega_{\rm s}^{3}\overline{\Omega_{\rm pr}}^{\,2}
}
{
\gamma_{\rm s}\gamma_{\rm pr}^{4}
}
e^{5i\delta\omega t}.
\end{equation}
Here the superscript ``ab'' denotes the antibunching limit. At fixed drive
voltages, \(\Omega_{\rm pr,s}=\sqrt{\gamma_{\rm pr,s}}\varepsilon_{\rm pr,s}\),
the amplitudes of the \(+3\delta\omega\) and \(-5\delta\omega\) peaks are
proportional to \(\gamma_{\rm s}\), whereas the \(+5\delta\omega\) peak is
proportional to \(\gamma_{\rm s}^2\). This is the analytical signature of the
antibunching-induced suppression of wave-mixing pathways involving several
source photons.

The peaks at \(\pm\delta\omega\) and \(-3\delta\omega\) are not suppressed in
the same way, because their leading pathways involve at most one source photon.
The suppression starts with side peaks whose leading monomials contain two or
more source-field factors, such as \(p_+s_+^2\) for \(+3\delta\omega\) and
\(p_-^3s_-^2\) for \(-5\delta\omega\).

\subsection{Total coherence decay rate} 
The two limiting regimes discussed above were formulated in terms of the radiative decay rates $\gamma_{\rm{s}}$ and $\gamma_{\rm{pr}}$, since the present model assumes radiatively limited qubits and neglects pure dephasing. More generally, however, the spectral comparison is determined by the total coherence decay rates, which set the linewidth
of the source radiation and the spectral response bandwidth of the probe,
respectively. These rates are
\begin{equation} \Gamma_{2,j}=\frac{\gamma_j}{2}+\gamma_{\phi,j}, \qquad j={\rm{s,pr}}, 
\end{equation} where $\gamma_{\phi,j}$ is the pure-dephasing rate. The coherent-filtering condition should therefore be understood more generally as $\Gamma_{2,\rm{s}}\gg\Gamma_{2,\rm{pr}}$, whereas the broadband-probe condition corresponds to $\Gamma_{2,\rm{pr}}\gg\Gamma_{2,\rm{s}}$. In the radiatively limited case considered throughout the analytical derivation, these conditions reduce to $\gamma_{\rm{s}}\gg\gamma_{\rm{pr}}$ and $\gamma_{\rm{pr}}\gg\gamma_{\rm{s}}$, respectively.

\section{Peak hierarchy and numerical comparison}
\label{sec:Discussion}

The weak-drive expansion provides a direct way to interpret the QWM side-peak
hierarchy. Each peak is associated with a definite monomial in the four drive
components \(p_\pm,s_\pm\), and the number of source-field factors in this
monomial determines whether the peak is sensitive to antibunching. The leading
monomials are summarized in Table~\ref{tab:monomials}.

\begin{table}[h!]
\caption{\label{tab:monomials}
Leading weak-drive monomials generating the first QWM side peaks of the probe.
The last column shows the small-\(\gamma_{\rm s}/\gamma_{\rm pr}\) scaling of
the normalized peak amplitude at fixed drive voltages.}
\begin{ruledtabular}
\begin{tabular}{cccc}
Peak & Leading monomial & Number of source factors & Broadband scaling \\
\hline
\(-\delta\omega\) & \(p_-\) & 0 & not suppressed \\
\(+\delta\omega\) & \(s_+\) & 1 & not suppressed \\
\(-3\delta\omega\) & \(p_-^2s_-\) & 1 & not suppressed \\
\(+3\delta\omega\) & \(p_+s_+^2\) & 2 &
\(\gamma_{\rm s}/\gamma_{\rm pr}\) \\
\(-5\delta\omega\) & \(p_-^3s_-^2\) & 2 &
\((3/2)\gamma_{\rm s}/\gamma_{\rm pr}\) \\
\(+5\delta\omega\) & \(p_+^2s_+^3\) & 3 &
\((5/4)(\gamma_{\rm s}/\gamma_{\rm pr})^2\)
\end{tabular}
\end{ruledtabular}
\end{table}

\begin{figure}[h!]
\includegraphics[width=4cm, height=3cm]{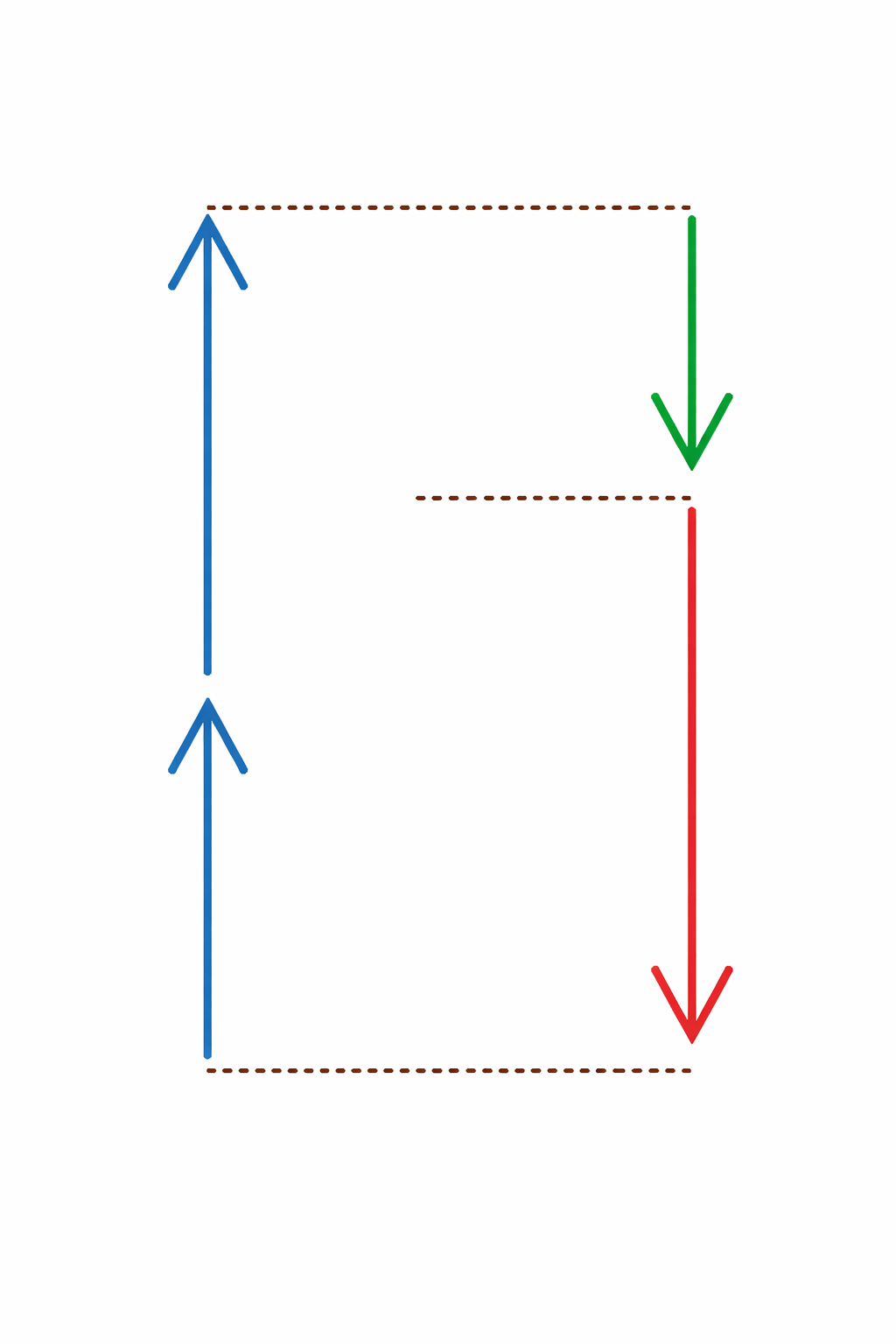}
\caption{\raggedright
Schematic representation of a multiphoton process generating the
\(-3\delta\omega\) side peak. Blue arrows indicate absorption of photons at the
probe drive frequency \(\omega_{\rm pr}\). Green arrows correspond to the
source frequency \(\omega_{\rm s}\). The red arrow denotes emission at the
mixed frequency \(2\omega_{\rm pr}-\omega_{\rm s}\), which corresponds to the
\(-3\delta\omega\) component in the rotating frame.}
\label{fig:arrows}
\end{figure}

This table gives a compact physical picture. The \(-3\delta\omega\) peak,
shown schematically in Fig.~\ref{fig:arrows}, is generated by a process
involving two photons of the probe tone and one source-frequency photon. Since
only one source photon is involved, the peak survives in the
antibunched regime. By contrast, the \(+3\delta\omega\), \(-5\delta\omega\),
and \(+5\delta\omega\) peaks require two or more source-field factors in their
leading pathways and are therefore suppressed when the source field is
antibunched.

To quantify the suppression, we use the voltage parametrization \ref{voltages def}
and compare the side-peak amplitudes with their coherent-filtering values at
the same \(\varepsilon_{\rm pr}\), \(\varepsilon_{\rm s}\), and
\(\gamma_{\rm pr}\). We define
\begin{equation}
S^{n\delta\omega}_{\rm{pr}}=|\braket{\sigma_{-}^{\rm{pr}}}_{+n\delta\omega}|/|\braket{\sigma^{\rm{pr}}_{-}}^{\rm{coh}}_{n\delta\omega}|
\end{equation}
where the denomenator corresponds to the coherent filtering limit \(\gamma_{\rm s}^{\rm{coh}}\rightarrow\infty\) and is defined by (\ref{coh+coh +-1 delta})-(\ref{coh+coh +-5 delta}) together with (\ref{effective Rabis})-(\ref{voltages def}). From
Eqs.~(\ref{+3 peak})--(\ref{+5 peak}) we obtain
\begin{equation}\label{S plus 3 full}
S_{+3}
=
\frac{\gamma_{\rm s}}
{\gamma_{\rm s}+\gamma_{\rm pr}},
\end{equation}
\begin{equation}\label{S minus 5 full}
S_{-5}
=
\frac{\gamma_{\rm s}
\left(\gamma_{\rm s}+\frac{3}{2}\gamma_{\rm pr}\right)}
{(\gamma_{\rm s}+\gamma_{\rm pr})^2},
\end{equation}
and
\begin{equation}\label{S plus 5 full}
S_{+5}
=
\frac{\gamma_{\rm s}^2
\left(\gamma_{\rm s}+\frac{5}{2}\gamma_{\rm pr}\right)}
{
(\gamma_{\rm s}+\gamma_{\rm pr})^2
(\gamma_{\rm s}+2\gamma_{\rm pr})
}.
\end{equation}
In the antibunched limit ($\gamma_{\rm{s}}\ll\gamma_{\rm{pr}}$) these expressions reduce to
\begin{equation}\label{ratio +3}
S_{+3}\simeq
\frac{\gamma_{\rm s}}{\gamma_{\rm pr}},
\end{equation}
\begin{equation}\label{ratio -5}
S_{-5}\simeq
\frac{3}{2}\frac{\gamma_{\rm s}}{\gamma_{\rm pr}},
\end{equation}
\begin{equation}\label{ratio +5}
S_{+5}\simeq
\frac{5}{4}
\left(\frac{\gamma_{\rm s}}{\gamma_{\rm pr}}\right)^2.
\end{equation}
Thus the side peaks involving two source photons are linearly suppressed in
\(\gamma_{\rm s}/\gamma_{\rm pr}\), whereas the peak involving three source
photons is quadratically suppressed.

We now compare the analytical formulas with numerical solutions of the
stationary cascaded equations. The numerical calculation solves
Eqs.~(\ref{sigma_m^p})--(\ref{sigma_zm}) without expanding in the drive
amplitudes, while using the same stationary approximation. The Fourier
components of \(\braket{\sigma_-^{\rm pr}}\) are then extracted from the
resulting quasiperiodic solution.

\begin{figure}[h!]
    \centering
    \begin{subfigure}[b]{0.34\textwidth}
        \centering
        \includegraphics[width=\textwidth, height=45mm]{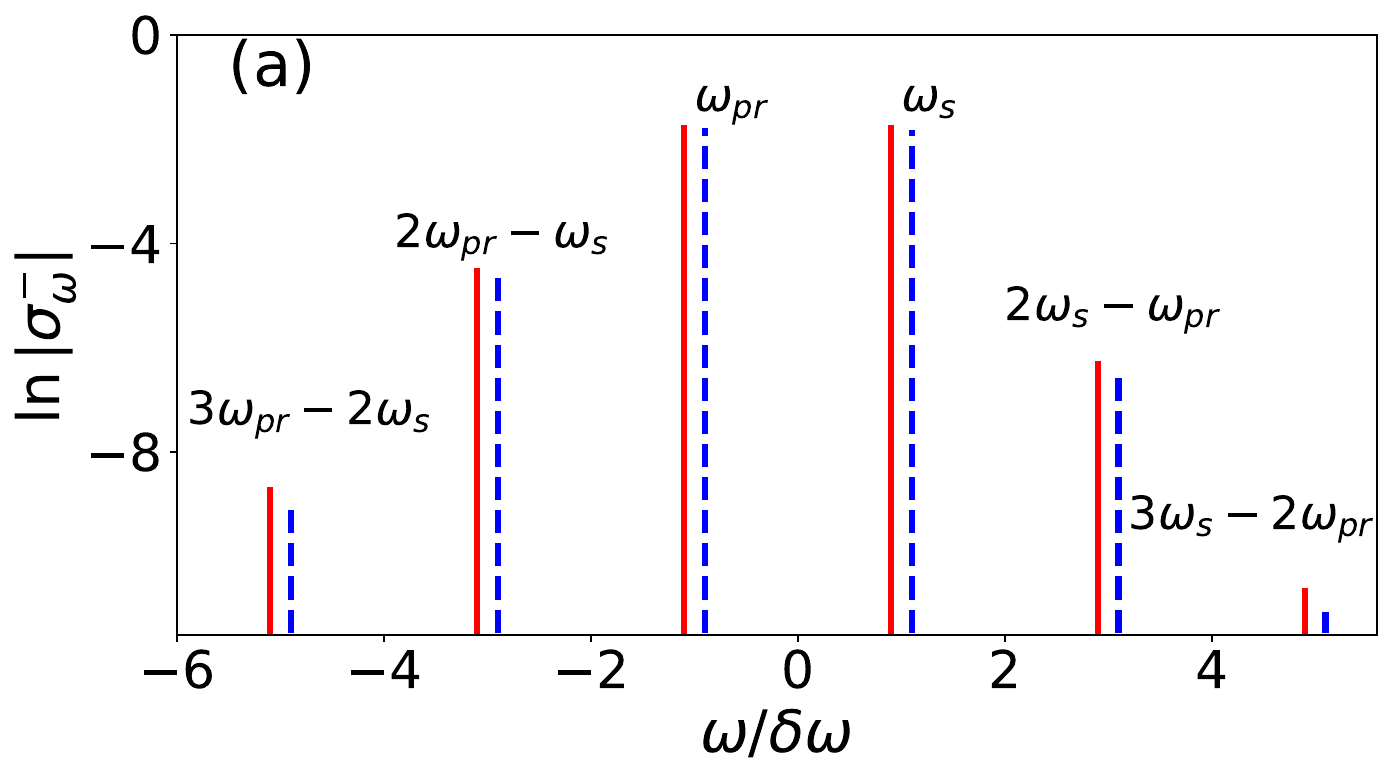}
        \label{fig1:a}
    \end{subfigure}
    \begin{subfigure}[b]{0.3\textwidth}
        \centering
        \includegraphics[width=\textwidth, height=44mm]{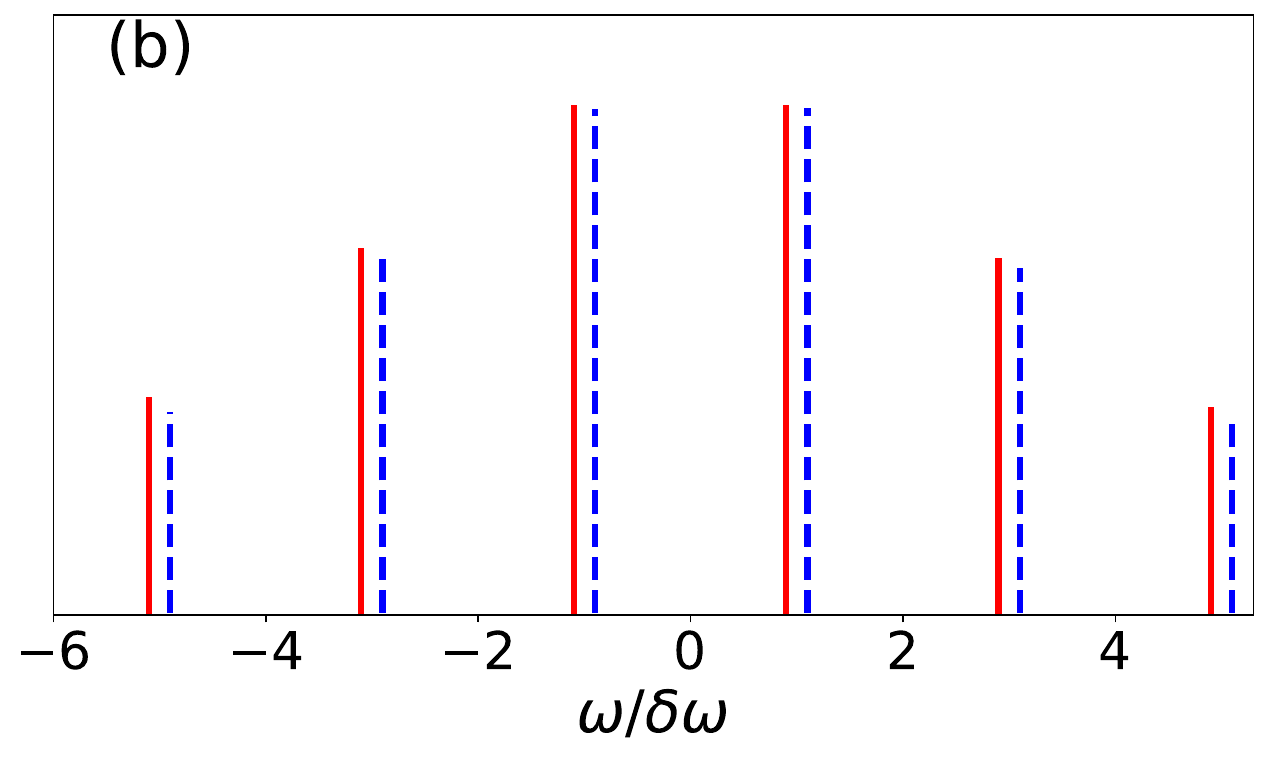}
        \label{fig1:b}
    \end{subfigure}
    \begin{subfigure}[b]{0.3\textwidth}
        \centering
        \includegraphics[width=\textwidth, height=44mm]{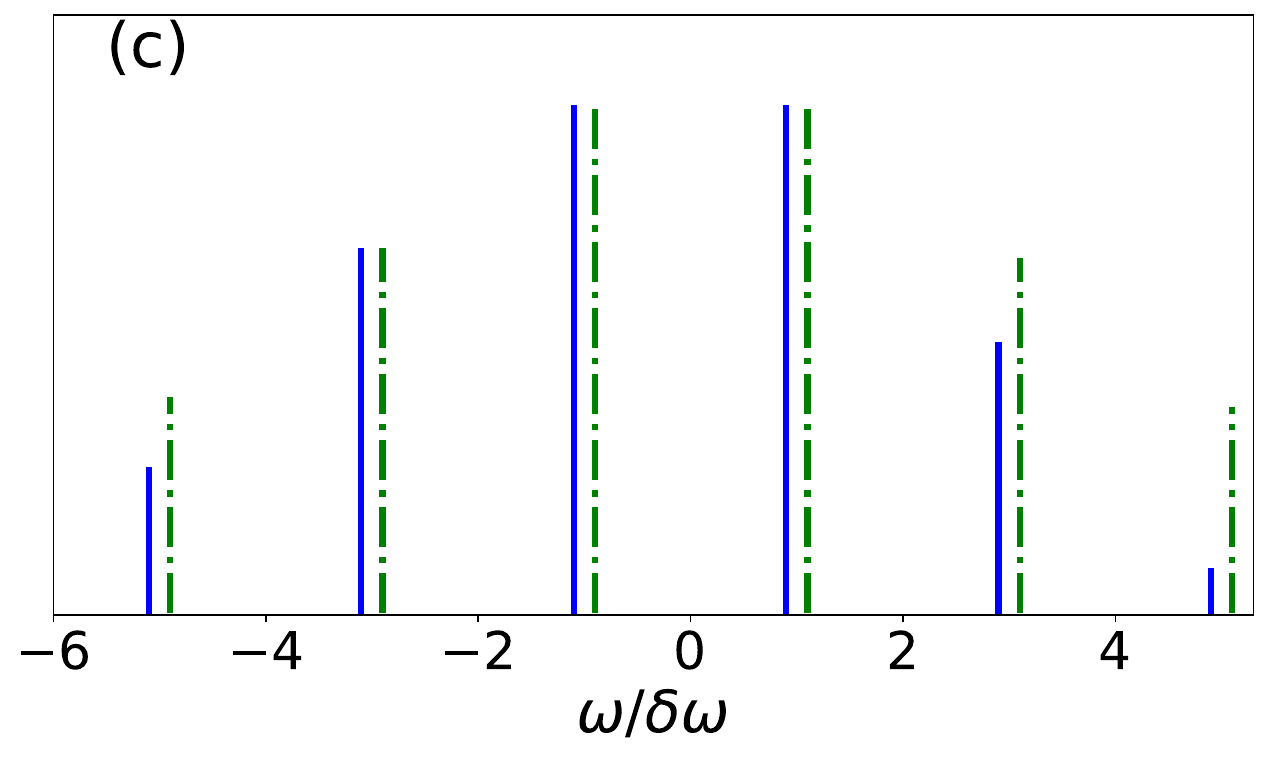}
        \label{fig1:c}
    \end{subfigure}
\caption{\raggedright
(a) QWM spectrum in the antibunching regime. The analytical result
(red solid line) is compared with the numerical solution of the stationary
cascaded equations (blue dashed line). (b) The same comparison in the
coherent-filtering regime. (c) Analytical comparison of the
antibunching spectrum (blue solid line) and the coherent-filtering spectrum
(green dash-dotted line). In both cases
\(\gamma_{\rm pr}=5\), \(\varepsilon_{\rm pr}=0.2\), and
\(\varepsilon_{\rm s}=0.1\). In the coherent-filtering case
\(\gamma_{\rm s}=25\), whereas in the antibunching case
\(\gamma_{\rm s}=1\).}
\label{fig:quantum classical comparsion}
\end{figure}

Figure~\ref{fig:quantum classical comparsion} shows that the peaks at
\(\pm\delta\omega\) and \(-3\delta\omega\) are the same in the two
limits, whereas the \(+3\delta\omega\), \(-5\delta\omega\), and
\(+5\delta\omega\) peaks are strongly suppressed in the antibunching
regime. This is consistent with the photon statistics of resonance fluorescence:
the probability of emitting two or more photons within a short time interval is
suppressed for a two-level source \cite{filtered_statistics}. The frequency
domain peak hierarchy therefore provides a spectroscopic fingerprint of this
antibunching.

\begin{figure*}[t]
    \centering

    \begin{subfigure}[b]{0.47\textwidth}
        \centering
        \includegraphics[width=\textwidth]{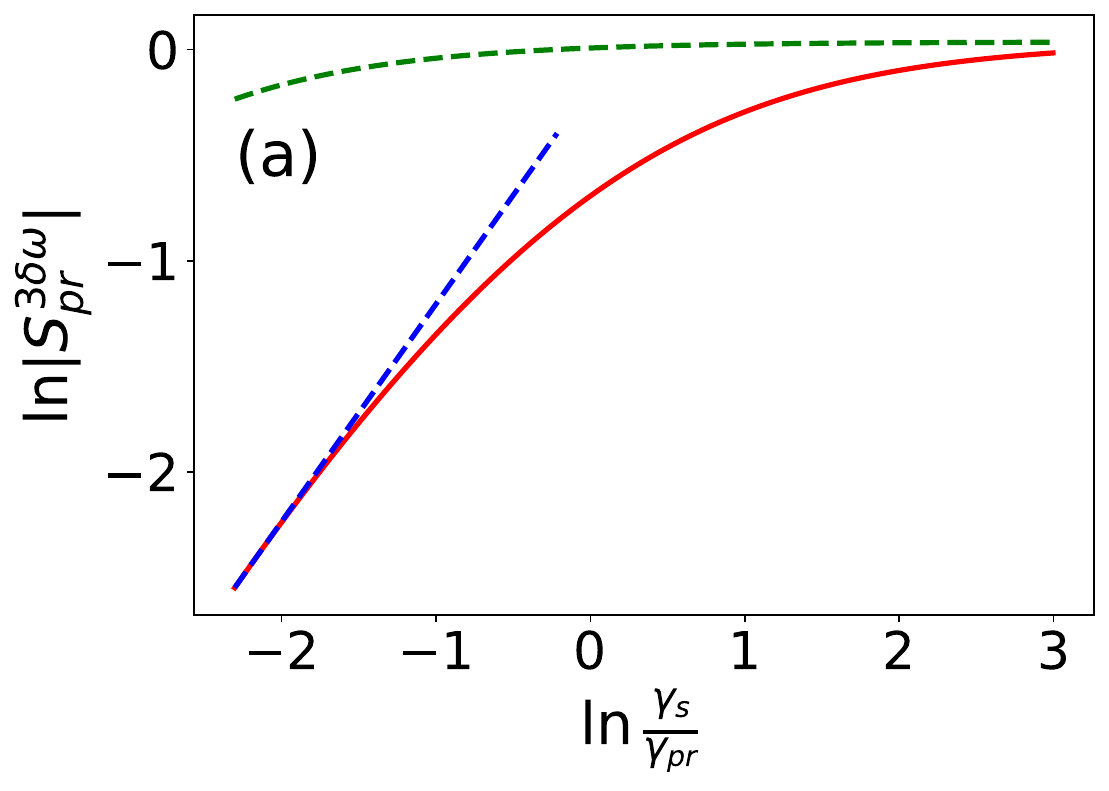}
        \label{fig:gamma_peak3}
    \end{subfigure}
    \hfill
    \begin{subfigure}[b]{0.47\textwidth}
        \centering
        \includegraphics[width=\textwidth]{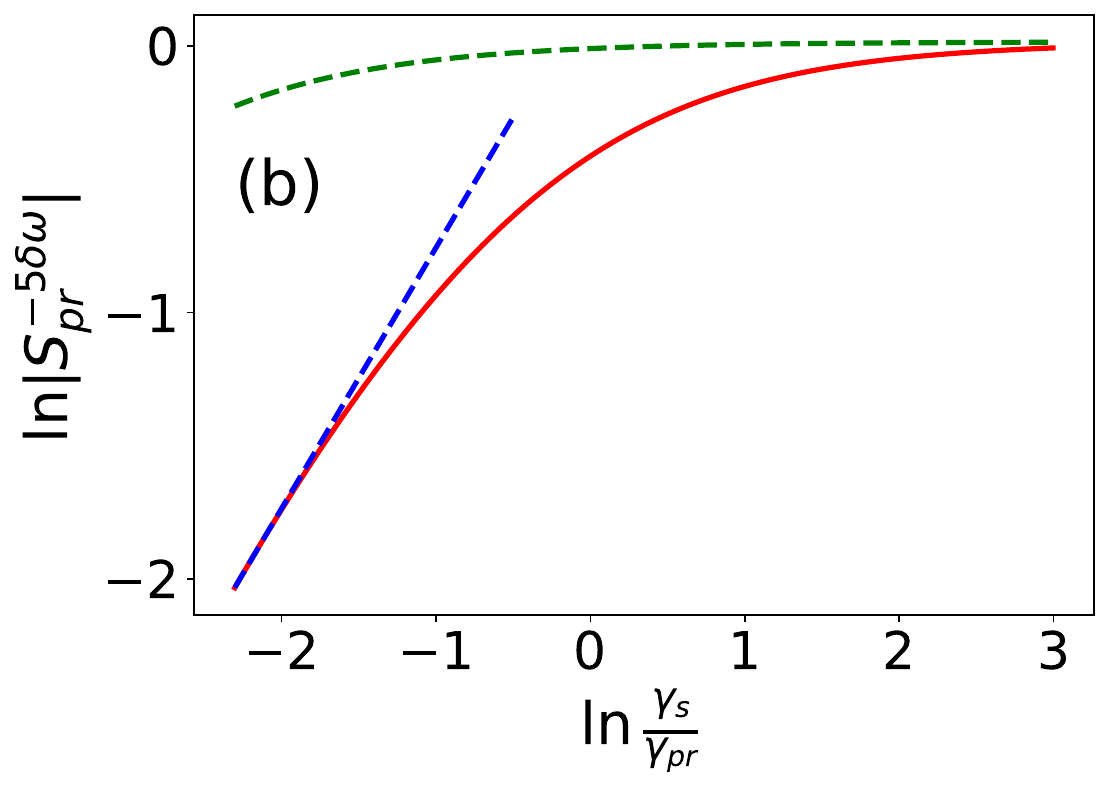}
        \label{fig:gamma_peakmin5}
    \end{subfigure}

    \vspace{0.4cm}

    \begin{subfigure}[b]{0.47\textwidth}
        \centering
        \includegraphics[width=\textwidth]{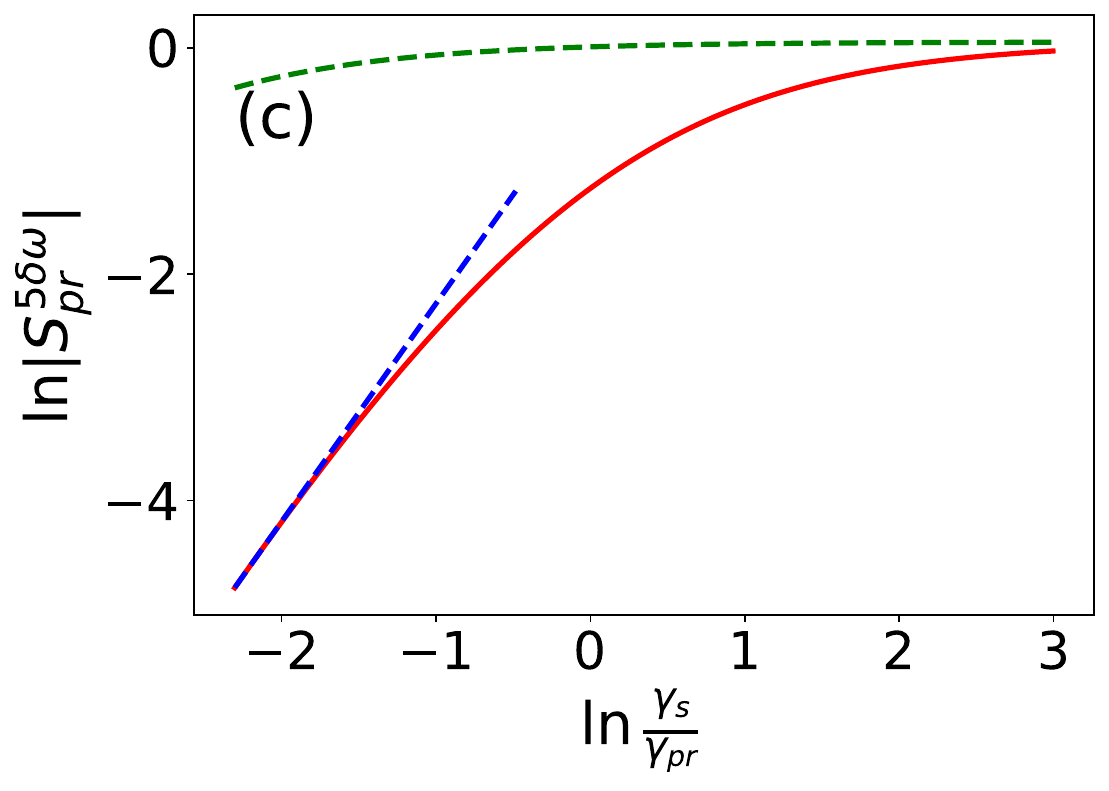}
        \label{fig:gamma_peak5}
    \end{subfigure}
    \hfill
    \begin{subfigure}[b]{0.47\textwidth}
        \centering
        \includegraphics[width=\textwidth]{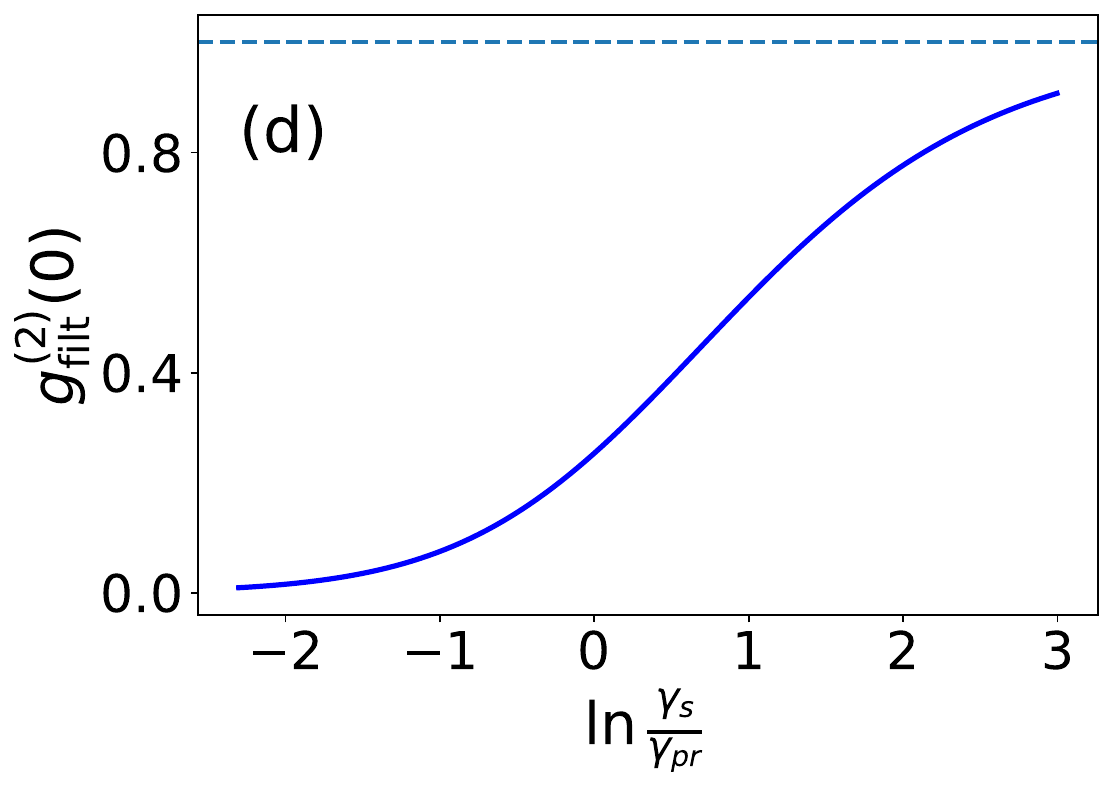}
        \label{fig:gamma_g2}
    \end{subfigure}

    \caption{\raggedright
Panels (a)--(c) show the dependence of the normalized side-peak amplitudes on the source--probe linewidth ratio:
(a) \(+3\delta\omega\), (b) \(-5\delta\omega\), and
(c) \(+5\delta\omega\). Red solid lines show numerical results obtained from
the stationary cascaded equations. Blue dashed lines show the antibunching
asymptotes, and green dashed lines
show the factorized mean-field solution obtained by neglecting
source--probe correlations while keeping the same system parameters and
fixed-voltage driving.
The normalized amplitude is 
$S^{n\delta\omega}_{\rm{pr}}=\braket{\sigma_{-}^{\rm{pr}}}_{n\delta\omega}/\braket{\sigma^{\rm{pr}}_{-}}^{\rm{coh}}_{n\delta\omega}$, where the denominator is the corresponding coherent-filtering value in
the limit $\gamma_{\rm s}/\gamma_{\rm{pr}}\rightarrow\infty$. Panel (d) shows the zero-delay second-order correlation function
$g_{\rm filt}^{(2)}(0)$ of the source radiation after Lorentzian
bandwidth filtering with filter linewidth
$\kappa=\gamma_{\rm pr}$. The horizontal dashed line marks
$g_{\rm filt}^{(2)}(0)=1$, corresponding to coherent-field statistics.}
\label{fig:gamma dependence}
\end{figure*}

Figure~\ref{fig:gamma dependence} provides two additional tests of the
origin of the sideband suppression. First, we compare the full cascaded
solution with a factorized mean-field approximation. In the latter, all
mixed source--probe correlators appearing in Eqs.~(3)--(9) are replaced
by products of single-qubit averages,
\begin{equation}
    \langle A_{\rm s} B_{\rm pr}\rangle
    \rightarrow
    \langle A_{\rm s}\rangle
    \langle B_{\rm pr}\rangle .
\end{equation}
For example,
$\langle\sigma_-^{\rm s}\sigma_z^{\rm pr}\rangle
\rightarrow
\langle\sigma_-^{\rm s}\rangle
\langle\sigma_z^{\rm pr}\rangle$.
This approximation retains the exact stationary mean field of the
source, including its dependence on $\gamma_{\rm s}$ and its saturation,
but removes the source--probe quantum correlations. The corresponding
mean-field side-peak amplitudes are shown by the green dashed curves in
Figs.~\ref{fig:gamma dependence}(a)--(c). They remain close to the
coherent-wave-mixing values over the range where the full cascaded
solution exhibits a strong suppression. Thus, the decrease of the
sidebands cannot be accounted for by the linewidth dependence of the
mean source field alone.

Second, to characterize the photon statistics of the part of the source
radiation sampled within the probe bandwidth, we introduce a
bandwidth-filtered source field
\begin{equation}
    a_{\rm f}(t)
    =
    \int_0^\infty d\tau\,
    h_{\rm pr}(\tau)\,
    a_{\rm out,s}(t-\tau),
\end{equation}
where $h_{\rm pr}(\tau)\propto
\exp(-\gamma_{\rm pr}\tau/2)$ is a Lorentzian filter response with the
same characteristic bandwidth as the probe. We then evaluate
\begin{equation}
    g_{\rm filt}^{(2)}(0)
    =
    \frac{
    \langle
    a_{\rm f}^\dagger a_{\rm f}^\dagger
    a_{\rm f}a_{\rm f}
    \rangle
    }{
    \langle a_{\rm f}^\dagger a_{\rm f}\rangle^2
    }.
\end{equation}
The result is shown in Fig.~\ref{fig:gamma dependence}(d). In the
broadband-probe regime,
$\gamma_{\rm pr}\gg\gamma_{\rm s}$,
the filtered field retains the antibunching of the source fluorescence,
and $g_{\rm filt}^{(2)}(0)$ approaches zero. As
$\gamma_{\rm s}/\gamma_{\rm pr}$ increases, the narrower probe bandwidth
selects an increasingly coherent component of the source radiation and
$g_{\rm filt}^{(2)}(0)$ approaches unity. The crossover of the filtered
photon statistics occurs on the same linewidth scale as the suppression
of the multiphoton QWM sidebands.

For small \(\gamma_{\rm s}/\gamma_{\rm pr}\), the
\(+3\delta\omega\) and \(-5\delta\omega\) peaks follow the linear laws
(\ref{ratio +3}) and (\ref{ratio -5}), while the \(+5\delta\omega\) peak
follows the quadratic law (\ref{ratio +5}). For large
\(\gamma_{\rm s}/\gamma_{\rm pr}\), all three normalized amplitudes saturate to
unity, as expected in the coherent-filtering regime. The numerical results
therefore confirm both the analytical side-peak formulas and the physical
interpretation in terms of antibunching-induced suppression of multiphoton
source pathways.

The small $\gamma_{\rm{s}}/\gamma_{\rm{pr}}$ part of
Fig.~\ref{fig:gamma dependence} should be interpreted within the
radiatively limited model used in the present calculation. In the
presence of source pure dephasing, the source coherence linewidth is
$\Gamma_{2,\rm{s}}=\gamma_{\rm{s}}/2+\gamma_{\phi,\rm{s}}$.
Consequently, decreasing $\gamma_{\rm{s}}$ produces a proportional
narrowing of the source fluorescence spectrum only while
$\gamma_{\phi,\rm{s}}\ll\gamma_{\rm{s}}/2$. Once the radiative and
pure-dephasing contributions become comparable,
$\gamma_{\rm{s}}/\gamma_{\rm{pr}}$ no longer represents the ratio of
the physical source and probe linewidths. The small
$\gamma_{\rm{s}}/\gamma_{\rm{pr}}$ asymptotes in
Eqs.~(\ref{ratio +3})--(\ref{ratio +5}) should therefore not be
extrapolated into this regime without explicitly including pure
dephasing in the cascaded equations.
For the device of Ref.~\cite{paper2},
$\gamma_{\rm{s}}/(2\pi)=1.74~\mathrm{MHz}$ and
$\gamma_{\phi,\rm{s}}/(2\pi)=0.15~\mathrm{MHz}$, so that the radiative
contribution $\gamma_{\rm{s}}/2$ exceeded the pure-dephasing
contribution by a factor of approximately $5.8$. At an unchanged
pure-dephasing rate, reducing $\gamma_{\rm{s}}$ by approximately the
same factor would already bring the source to
$\gamma_{\rm{s}}/2\simeq\gamma_{\phi,\rm{s}}$. Thus, the regime
$\gamma_{\rm{pr}}\gg\gamma_{\rm{s}}$ cannot be approached arbitrarily
far in this device by decreasing $\gamma_{\rm{s}}$ alone.
Moreover, in the implementation of Ref.~\cite{paper2}, the radiative
coupling rates were set by the device design and could not be varied
independently after fabrication.

The parameters of the experimental cascaded device in
Ref.~\cite{paper2} also provide a direct estimate of the observability
of the predicted suppression. For
$\gamma_{\rm{s}}/(2\pi)\simeq1.7~\mathrm{MHz}$ and
$\gamma_{\rm{pr}}/(2\pi)\simeq1.8~\mathrm{MHz}$, one has
$\gamma_{\rm{s}}/\gamma_{\rm{pr}}\simeq0.94$. Substitution into
Eqs.~(\ref{S plus 3 full})--(\ref{S plus 5 full}) gives
$S_{+3}\simeq0.486$, $S_{-5}\simeq0.611$, and
$S_{+5}\simeq0.276$. These amplitude ratios correspond to relative
peak-power suppressions of approximately $6.3$, $4.3$, and $11.2$ dB,
respectively. In Ref.~\cite{paper2}, contrasts of $7.1$ dB for the
$+3\delta\omega$ component and $4.2$ dB for the
$-5\delta\omega$ component were resolved against a matched coherent
reference using a spectrum-analyzer resolution bandwidth of $5$ Hz and
a source-to-probe field-amplitude transmission of approximately $0.79$. Characteristic Rabi drive amplitudes in the experiment of
Ref.~\cite{paper2} were of the order of tenths to units of megahertz.

\section{Conclusion}
\label{sec:Conclusion}

We have developed an analytical theory of quantum wave mixing in a cascaded
source--probe qubit system. The central technical step is a weak-drive Taylor
expansion of the stationary cascaded response. The decay rates and the
unidirectional source--probe coupling are kept in the pump-independent matrix
\(\hat A\), while the coherent drives enter through a drive matrix
\(\hat\Omega\). Expanding \((\hat A+\hat\Omega)^{-1}\) as a Neumann series gives
a systematic and controlled hierarchy of contributions to the probe coherence.

This formulation yields closed-form expressions for the leading QWM side peaks
at \(\pm\delta\omega\), \(\pm3\delta\omega\), and
\(\pm5\delta\omega\). The formulas identify the drive monomial responsible for
each peak and reveal its dependence on the cascaded coupling and on the
linewidth ratio \(\gamma_{\rm s}/\gamma_{\rm pr}\). In the
coherent-filtering limit, \(\gamma_{\rm s}\gg\gamma_{\rm pr}\), the source
reduces to an effective coherent tone and the known coherent--coherent QWM
hierarchy is recovered \cite{Dmitriev2019,paper1}. In the opposite antibunching
limit, \(\gamma_{\rm pr}\gg\gamma_{\rm s}\), the probe is sensitive to the
antibunched resonance fluorescence of the source. Side peaks whose leading
pathways contain two or more source-field factors are then parametrically
suppressed.

The resulting scaling laws provide a direct frequency-domain signature of
antibunched itinerant radiation. Peaks involving one source photon, such as
\(-3\delta\omega\), remain unsuppressed, while peaks involving two
or three source photons are reduced by powers of
\(\gamma_{\rm s}/\gamma_{\rm pr}\). Numerical solutions of the stationary
cascaded equations agree with these analytical predictions and confirm the
crossover between the antibunching and coherent-filtering regimes.

The present work also clarifies its relation to earlier studies of the
same cascaded architecture. Ref.~\cite{paper2} demonstrated the
sideband suppression experimentally and reproduced it using numerical
solutions of the full cascaded dynamics. The new contribution here is
a closed-form weak-drive theory that assigns each side peak to a
definite drive monomial and provides explicit crossover functions and
asymptotic scaling laws in
$\gamma_{\rm{s}}/\gamma_{\rm{pr}}$. Ref.~\cite{paper3}, in contrast,
treats the complementary strong-drive regime in which a narrow probe
selects the pair-correlated central Mollow component and the QWM
spectrum obeys an odd-source-field parity rule. The hierarchy obtained
here has a different origin: it follows from antibunching and
parametrically suppresses all leading pathways containing two or more
source-field factors. Taken together, these results show how the same
source--probe circuit connects coherent filtering,
antibunching-dominated multiphoton suppression, and pair-correlated QWM
as the source excitation and the probe bandwidth are varied.

More broadly, the theory shows that cascaded QWM can be used not only as a
probe of the nonlinear response of a superconducting qubit, but also as a
spectroscopic diagnostic of the photon statistics of the field incident on it.
The analytical framework developed here provides a compact basis for
interpreting peak suppression in cascaded QWM experiments and for extending
wave-mixing spectroscopy to other forms of nonclassical microwave radiation
\cite{Dmitriev2017,Decrinis2018,Dmitriev2019,paper1,paper2}.

\section{Acknowledgements}
The study is supported by the Ministry of Science and Higher Education of the Russian Federation (agreement No. 075-15-2024-538).

\appendix

\section{Matrix form of the stationary cascaded equations}
\label{app:matrices}

In this Appendix we give the explicit matrices used in
Sec.~\ref{sec:perturbations}. We use the ordering of variables defined in
Eq.~(\ref{vector X}). The vector of variables
\[
\begin{split}
\vec X=&
\bigl(
\braket{\sigma_{-}^{\rm{pr}}},
\braket{\sigma_{-}^{\rm{s}}\sigma_{z}^{\rm{pr}}},
\braket{\sigma_{z}^{\rm{s}}\sigma_{-}^{\rm{pr}}},
\braket{\sigma_{+}^{\rm{pr}}},
\braket{\sigma_{+}^{\rm{s}}\sigma_{z}^{\rm{pr}}},
\braket{\sigma_{z}^{\rm{s}}\sigma_{+}^{\rm{pr}}},
\\
&\braket{\sigma_{-}^{\rm{s}}\sigma_{-}^{\rm{pr}}},
\braket{\sigma_{+}^{\rm{s}}\sigma_{+}^{\rm{pr}}},
\braket{\sigma_{z}^{\rm{pr}}},
\braket{\sigma_{+}^{\rm{s}}\sigma_{-}^{\rm{pr}}},
\braket{\sigma_{-}^{\rm{s}}\sigma_{+}^{\rm{pr}}},
\braket{\sigma_{z}^{\rm{s}}\sigma_{z}^{\rm{pr}}}
\bigr)^T .
\end{split}
\]

Introduce the abbreviations
\[
r=\frac{\gamma_{\rm s}}{\gamma_{\rm pr}},
\qquad
p_-=\frac{\Omega_{\rm pr}}{\gamma_{\rm pr}}e^{-i\delta\omega t},
\qquad
p_+=\frac{\overline{\Omega_{\rm pr}}}{\gamma_{\rm pr}}e^{i\delta\omega t},
\]
\[
s_+=\frac{\Omega_{\rm s}}{\gamma_{\rm pr}}e^{i\delta\omega t},
\qquad
s_-=\frac{\overline{\Omega_{\rm s}}}{\gamma_{\rm pr}}e^{-i\delta\omega t}.
\]
The pump-independent matrix \(\hat A\) is block diagonal in the chosen ordering,
\[
\hat A=A_-\oplus A_+\oplus A_{--}\oplus A_{++}\oplus A_z .
\]
The first block acts on
\((\braket{\sigma_-^{\rm pr}},
\braket{\sigma_-^{\rm s}\sigma_z^{\rm pr}},
\braket{\sigma_z^{\rm s}\sigma_-^{\rm pr}})\):
\[
A_-=
\begin{pmatrix}
-\frac12 & \alpha & 0\\
-\alpha & -\left(1+\frac r2\right) & -\alpha\\
-r & -\alpha & -\left(r+\frac12\right)
\end{pmatrix}.
\]
The block \(A_+\), acting on
\((\braket{\sigma_+^{\rm pr}},
\braket{\sigma_+^{\rm s}\sigma_z^{\rm pr}},
\braket{\sigma_z^{\rm s}\sigma_+^{\rm pr}})\), has the same form,
\[
A_+=
\begin{pmatrix}
-\frac12 & \alpha & 0\\
-\alpha & -\left(1+\frac r2\right) & -\alpha\\
-r & -\alpha & -\left(r+\frac12\right)
\end{pmatrix}.
\]
For the two variables
\(\braket{\sigma_-^{\rm s}\sigma_-^{\rm pr}}\) and
\(\braket{\sigma_+^{\rm s}\sigma_+^{\rm pr}}\), one has
\[
A_{--}=A_{++}=-\frac{r+1}{2}.
\]
Finally, the block acting on
\((\braket{\sigma_z^{\rm pr}},
\braket{\sigma_+^{\rm s}\sigma_-^{\rm pr}},
\braket{\sigma_-^{\rm s}\sigma_+^{\rm pr}},
\braket{\sigma_z^{\rm s}\sigma_z^{\rm pr}})\) is
\[
A_z=
\begin{pmatrix}
-1 & -2\alpha & -2\alpha & 0\\
\frac{\alpha}{2} & -\frac{r+1}{2} & 0 & \frac{\alpha}{2}\\
\frac{\alpha}{2} & 0 & -\frac{r+1}{2} & \frac{\alpha}{2}\\
-r & 2\alpha & 2\alpha & -(r+1)
\end{pmatrix}.
\]
These blocks are nonsingular for \(r>0\). In particular,
\[
\det A_-=\det A_+=-\frac{(r+2)(2r+1)}{8},
\qquad
\det A_z=\frac{(r+1)^3}{4}.
\]

The drive-dependent matrix \(\hat\Omega\) is linear in
\(p_\pm,s_\pm\) and reads

\[
\hat{\Omega}=\left[\begin{array}{cccccccccccc}0 & 0 & 0 & 0 & 0 & 0 & 0 & 0 & p_{-} & 0 & 0 & 0\\0 & 0 & 0 & 0 & 0 & 0 & - 2 p_{+} & 0 & 0 & 0 & - 2 p_{-} & s_{+}\\0 & 0 & 0 & 0 & 0 & 0 & - 2 s_{-} & 0 & 0 & - 2 s_{+} & 0 & p_{-}\\0 & 0 & 0 & 0 & 0 & 0 & 0 & 0 & p_{+} & 0 & 0 & 0\\0 & 0 & 0 & 0 & 0 & 0 & 0 & - 2 p_{-} & 0 & - 2 p_{+} & 0 & s_{-}\\0 & 0 & 0 & 0 & 0 & 0 & 0 & - 2 s_{+} & 0 & 0 & - 2 s_{-} & p_{+}\\0 & p_{-} & s_{+} & 0 & 0 & 0 & 0 & 0 & 0 & 0 & 0 & 0\\0 & 0 & 0 & 0 & p_{+} & s_{-} & 0 & 0 & 0 & 0 & 0 & 0\\- 2 p_{+} & 0 & 0 & - 2 p_{-} & 0 & 0 & 0 & 0 & 0 & 0 & 0 & 0\\0 & 0 & s_{-} & 0 & p_{-} & 0 & 0 & 0 & 0 & 0 & 0 & 0\\0 & p_{+} & 0 & 0 & 0 & s_{+} & 0 & 0 & 0 & 0 & 0 & 0\\0 & - 2 s_{-} & - 2 p_{+} & 0 & - 2 s_{+} & - 2 p_{-} & 0 & 0 & 0 & 0 & 0 & 0\end{array}\right]
\]

The inhomogeneous vector is
\[
\vec b=
\begin{pmatrix}
0\\
\frac{2s_+}{r}F_s\\
0\\
0\\
\frac{2s_-}{r}F_s\\
0\\
0\\
0\\
-1\\
0\\
0\\
F_s
\end{pmatrix},
\qquad
F_s=\frac{1}{1+8s_+s_-/r^2}.
\]
Expanding \(F_s\) in powers of the source drive gives
\[
F_s=
1-\frac{8s_+s_-}{r^2}
+\frac{64s_+^2s_-^2}{r^4}
+O(|s|^6),
\]
which determines the vectors \(\vec b^{[N]}\) used in the recursion
(\ref{recursion}).

For reference, the sideband-producing fifth-order terms in the first component
of \(\vec X\) are
\[
X_{1,-5}^{[5]}
=
-\frac{256\alpha^2(2r+3)}
{r(r+1)^2}
p_-^3s_-^2,
\]
and
\[
X_{1,+5}^{[5]}
=
\frac{
512\alpha^3(2r+5)
}
{r(r+1)^2(r+2)}
p_+^2s_+^3 .
\]
These terms give Eqs.~(\ref{-5 peak}) and (\ref{+5 peak}) after returning to
the physical variables.


\end{document}